\documentclass[a4paper,11pt]{article}
\pdfoutput=1
\usepackage{amssymb}
\usepackage[utf8]{inputenc}
\usepackage{textcomp}
\usepackage{latexsym}
\usepackage{epstopdf}
\usepackage{epsfig}
\usepackage{graphicx}
\usepackage{amsmath}
\usepackage{multirow}
\usepackage{subfigure}
\usepackage{a4wide}
\usepackage{cite}
\usepackage{rotating}
\usepackage{hyperref}
\newcommand{\bwt}{\begin{widetext}}
\newcommand{\ewt}{\end{widetext}}
\newcommand{\be}{\begin{equation}}
\newcommand{\ee}{\end{equation}}
 \newcommand{\bdm}{\begin{displaymath}}
\newcommand{\edm}{\end{displaymath}}
\newcommand{\bea}{\begin{eqnarray}}
\newcommand{\eea}{\end{eqnarray}}
\newcommand{\nn}{\nonumber}

\usepackage{slashed}
\newcommand{\mathsym}[1]{{}}

\newcommand{\baz}{\begin{array}{cc}}
\newcommand{\bad}{\begin{array}{ccc}}
\newcommand{\ba}{\begin{array}{c}}
\newcommand{\ea}{\end{array}}
\newcommand{\bi}{\begin{itemize}}
\newcommand{\ei}{\end{itemize}}
\newcommand{\bmt}{\begin{pmatrix}}
\newcommand{\emt}{\end{pmatrix}}
\newcommand{\bt}{\begin{tabular}}
\newcommand{\et}{\end{tabular}}

\newcommand{\benu}{\begin{enumerate}}
\newcommand{\eenu}{\end{enumerate}}

\begin{document}

\title{\Large\bf {High Scale Type-II Seesaw, Dominant Double Beta Decay  within 
Cosmological Bound  and  LFV Decays in SU(5)}}

\author{M. K. Parida$^{\dagger}$, Rajesh Satpathy $^{*}$\\
Centre for Excllence in Theoretical and Mathematical Sciences\\
Siksha 'O' Anusandhan Deemed to be University (SOADU), Khandagiri
Square,\\
 Bhubaneswar 751030, India}
\maketitle

\begin{abstract}
 Very recently  a novel implementation of type-II seesaw mechanism for
neutrino mass has been proposed in SU(5) grand unified theory with a
number of desirable new physical phenomena beyond the standard model.  
Introducing heavy right-handed neutrinos and extra fermion singlets, in this
work we show how the type-I seeaw cancellation mechanism works in this
SU(5) framework.  Besides predicting verifiable
LFV decays,  we further  show that the model predicts
dominant double beta decay  with normal hierarchy or
inverted hierarchy of active light neutrino masses
in concordance with cosmological bound.  In addition a novel
right-handed neutrino mass generation mechanism, independent of type-II seesaw
predicted mass hierarchy, is suggested in this work.
\end{abstract}
\noindent{${}^{\dagger}$email:minaparida@soa.ac.in}\\
\noindent{${}^{*}$email:satpathy.rajesh.rajesh@gmail.com} 

\newpage 

\section{Introduction}\label{sec:intr}
Renormalisable  standard model (SM) predicts  neutrinos to be massless 
 whereas  oscillation experiments prove them
to be massive \cite{Salas:2017,schwetz,forero,fogli,gonzalez}. All the generational mixings have been found to be much larger than the corresponding quark mixings.
Theoretically \cite{Alta:2014,AYS:2014,RNM:2015,RNM:2016,Valle:2016,Valle:2017,gs:2011,nurev:mkpbpn}
 neutrino masses are predicted through
various seesaw
mechanisms\cite{Minkowski:1977,Yanagida:1979,Gell-Mann:1979,Glashow:1979,RNM-gs:1980,Valle:1980,Valle:t2,Magg:t2,Lazaridis:t2,RNM-gs:t2,RNM-mkp:t2,inv1,inv2,inv3,inv4,inv5,inv6,inv7,LG:2000,type-III,Akhmedov-a,Akhmedov-b,mkp:hybrid,Linear-a,Linear-b,Linear-c,Rad,mkprad:2011,mkprad:2012}.
 In a minimal left-right
symmetric \cite{JCP:1974,rnmpati:1975,gs-RNM:1975,gs:1979} grand unified theory (GUT)
like $SO(10)$ \cite{georgi:1974,FM:1975} where 
 parity (P) violation in weak interaction is explained along with
 fermion masses
 \cite{babu-rnm:1993,Joshipura,Altarelli-Blankenburg,Bertollini,mkp-PRD:2008,Goh-Mohapatra}, a number
of these seesaw mechanisms can be naturally embedded while unifying
 the  three frorces of the SM
 \cite{cmp1:1984,cmp2:1984,cmp3:1985,cmp4-PLB:1984,mkp-PLB:1983,mkp-PRD:1983,mkp-pkp:1991,mkp-pkp:1992,mkp-PRD:1998,rnm-PLB:1992,Pal:1994,Shafi:1982,Shafi:1984,Berto:2009,mprs:2007,Jaydeep:2018,mkp-ARC:1982}. More recently
 precision gauge coupling unification has been successfully implemented in
 direct symmetry breaking of $SO(18) \to SM$ which may have high
 potential for new physics \cite{Wilczek-Valle}.

The SO(10) model that predicts the  most popular canonical seesaw as
well as the type-II seesaw has also the ability to explain baryon asymmetry of the universe via
leptogenesis through heavy RH neutrino \cite{fukuyana:1986} or Higgs
triplet  decays\cite{Hambye-gs-a,Hambye-gs-b,Hambye-gs-c}. But
because of underlying quark lepton symmetry \cite{JCP:1974}, the
type-I seesaw scale as well as RH$\nu$ masses are so large that the model
predicts negligible lepton flavor violating (LFV) decays like $\mu\to e\gamma \,$, $\tau\to
\mu\gamma \,$, $\tau\to e\gamma \,$, and $\mu \to e{\bar e}e$. Similarly
direct mediation of large mass of scalar triplet required for type-II
seesaw   gives negligible contribution to lepton number violating
(LNV) and lepton flavor violating (LFV) decays.
Ever since the proposal of left-right symmetry, extensive
investigations continue in search of experimentally observable double beta decay\cite{bbexpt1-Klapdoor,bbexpt2-a,bbexpt2-b,bbexpt3-a,bbexpt3-b} in the
$W_R-W_R$ channel \cite{gs:bb,mkp-bs:2015}. Adding new dimension to such lepton number violating (LNV) process, the
like-sign dilepton production  has been suggested as a
possible means of detection of $W_R$-boson at accelerator energies \cite{Keung-gs:1983},
particularly the LHC \cite{Bajc-gs:2007}. However, no such signals of TeV scale $W_R-$
 have been detected so far.
 Even if $W_R$ mass and seesaw scales are
large and  inaccessible for direct verification, neutrinoless
double beta decay ($0\nu 2\beta$) in the $W_L-W_L$ channel \cite{Valle:t2,Valle:bb1,Valle:nuD1,Valle:MajDM,Valle:nuD2,Valle:CPLRS,Valle:KeVDM} is predicted close to
observable limit with $\tau_{\beta \beta} \ge 10^{25}$ yrs provided
light neutrino masses predicted by such high-scale seesaw mechanisms
are  quasidegenerate (QD)  with each mass $m_{\nu} \ge {\cal O}(0.2)$
eV \cite{bbexpt1-Klapdoor} and their sum $ > 0.6$ eV.
But as noted by the recent Planck data such QD type masses   violate the
cosmological bound \cite{Planck15}
\be
\Sigma_{\nu}\equiv \Sigma_i {\hat m}_i\le 0.23~(\rm  eV). \label{eq:cbound}.
\ee
 The fact that such
QD type $\nu$ masses violate the cosmological bound may be unravelling another
basic fundamental reason  why  detection of double beta decay continues to
elude experimental observation since several decades. On the other hand, if
neutrinos have smaller NH or IH type masses, there is no hope for
detection of these LNV events in near future with RH$\nu$ extended SM.
In other words predicting observable double beta decay in the
$W_L-W_L$ channel with left-handed helicities of both the beta
particles has been a formidable problem confronting theoretical and
experimentral physicists. However, it has been shown that in case of dynamical seesaw mechanism generating Dirac neutrinos the seesaw scale is accessible for direct experimental verification \cite{Valle:dynamical}.

The path breaking discovery of  inverse seesaw \cite{inv1,inv2,inv3,inv4,inv5,inv6,inv7} with
one extra singlet fermion per generation not only opened up
the neutrino mass generation mechanism for direct experimental tests,
but also  lifted up 
lepton flavor violating (LFV) decays \cite{Valle:LFV1987} from the abysmal depth of
experimental inaccessibility of negligible branching ratios (
$ Br.(l_\alpha\to l_\beta \gamma) \sim 10^{-50}$) to the illuminating
salvation of profound observability ( Br.$\simeq
10^{-8}-10^{-16}$)
\cite{lfvexpt-a,lfvexpt-b,lfvexpt-c,lfvexpt-d,lfvexpt-e,lfvexpt-f}
which has been discussed extensively
\cite{ilakovac,Depp:2005,Arina:2008,Mal:2009,Hirsch:2009,Depp:2006,spdas:2012,antbnd,antu-a,antu-b,antu-c,non-unit,non-unit-a,non-unit-b,non-unit-c,non-unit-d,non-unit-e,non-unit-f,non-unit-g,non-unit-h}. Despite inverse seesaw,
 observable double beta decay in the $W_L-W_L$ channel and the non-QD type
 neutrino masses remained mutually
exclusive until both the RH neutrinos and singlet fermions ($S_i$) were
brought into the arena of LFV and LNV
conundrum through the much needed extension of the Higgs sector. The King-Kang
\cite{Kim-Kang:2006} mechanism cancelled out the ruling supremacy of canonical  
seesaw which was profoundly exploited in SO(10) models with the
introduction of both the SO(10) Higgs representations ${16}_H$ and ${126}^{\dagger}_H$ 
\cite{nurev:mkpbpn,mkp-bs:2015,app:2013,pp:2012,Majee:2009,mkp-ARC:2010,pas:2014,bpn-mkp:2015} with successful
prediction of  observable double beta decay in the
$W_L-W_L$ channel \cite{Valle:t2,Valle:bb1}. Very interestingly, even though high scale type-II seesaw
 can 
govern light neutrino masses of any hierarchy, possibility of observable LFV and double beta
decay prediction in the $W_L-W_L$ channel irrespective of light
neutrino mass hierarchies has been  realized at least
theoretically \cite{nurev:mkpbpn,bpn-mkp:2015}.

The purpose of this work is to point out that there are new
interesting physics realizations with suitable
extension of a non-SUSY SU(5) GUT model proposed recently \cite{scp:2018} where
type-II seesaw, precision coupling unification, verifiable proton
decay, scalar dark matter and vacuum stability have been already
predicted. However with naturally large type-II seesaw scale $> 10^{9.2}$ GeV, observable double beta
decay accessible to ongoing experiments
\cite{bbexpt1-Klapdoor,bbexpt2,bbexpt3} is possible in this model
too with QD type neutrinos only of
common mass
with $|m_0| \ge 0.2$ eV like many other high scale seesaw models as noted above.
In this work we 
make additional prediction that dominant double beta decay in the
$W_L-W_L$ channel  can be realized with NH or IH type
hierarchy consistent with much lighter neutrino masses $|m_{i}|<< 0.2$
eV. Thus, this realization is  consistent with
cosmological bound of eq.(\ref{eq:cbound}). Although such possibilities were realized earlier
in SO(10) with TeV scale $W_R$ or $Z^{\prime}$ bosons as noted above,
in SU(5) without the presence of left-right symmetry and associated
gauge bosons, we have shown here for the first time that 
 the dominant double beta decay is mediated by a
sterile neutrino (Majorana fermion singlet) of ${\cal O}(1)$ GeV mass
of first generartion. The model further predicts LFV decay branching
ratios only $4-5$ orders smaller than the current experimental limits.
An additional interesting part of the present work is the first suggestion of a new
mechanism for heavy RH$\nu$ mass generation that
permits these masses to have hierarchies independent of conventional
type-II seesaw prediction. Thus  highlights of the present model are
\begin{itemize}
\item{First implementation of type-I
  seesaw  cancellation  mechanism
 leading to the dominance of type-II seesaw in SU(5).}
\item{Prediction of verifiable LFV decays only $4-5$ orders smaller
  than the current experimental limits.}
\item{Prediction of dominant double beta decay in the $W_L-W_L$
  channel close to the current experimental limits for light neutrino
  masses of NH or IH type in concordance  with cosmological bound.}
\item{Suggestion  of a new right-handed neutrino mass
  generation mechanism independent of type-II predicted mass hierarchy.} 
\item{Precision gauge coupling unification with verifiable proton
  decay which is the same as discussed in \cite{scp:2018}.}
\end{itemize}

This paper is organised in the following manner.
In Sec.\ref{sec:model} we briefly review the SU(5) model along with gauge coupling
unification and predictions of the intermediate scales. In Sec.\ref{sec:cancel} we
discuss how type-I seesaw formula for active neutrino masses cancels out giving rise to dominance of
type-II seesaw and prediction of another type-I seesaw formula for sterile neutrino masses.
Fit to neutrino oscillation data is discussed in Sec.\ref{sec:t2fit}.
 In Sec.\ref{sec:rhnu} we suggest a new mechanism of RH$\nu$ mass generation. Prediction on
 LFV decay branching ratios is discussed in Sec.\ref{sec:lfv}.
Lifetime prediction for double beta decay is presented in Sec.\ref{sec:bb}.
In Sec.\ref{sec:sum}
we discuss the results of this work and state our conclusion.
 
\section{\bf A NON-SUPERSYMMETRIC SU(5) MODEL}\label{sec:model}
\subsection{Extension of SU(5)}
As noted in ref.\cite{kynshi-mkp:1993}, inclusion of
the scalar $\kappa(3,0,8)\subset {75}_H$ with mass
$M_{\kappa}=10^{9.23} {\rm GeV}$  in the extended non-SUSY SU(5)
achieves precision gauge coupling unification. Then it has been shown
in \cite{scp:2018} that
  type-II seesaw ansatz for neutrino mass is realized by inserting the
  entire Higgs multiplet ${15}_H\subset SU(5)$ containing the LH Higgs triplet
  $\Delta_L(3,-1, 1)$ at the same mass scale $ M_{15_H}=10^{12}{\rm GeV}$.
\bea
\kappa(3,0,8)&\subset&{75}_H,\, M_{\kappa}=10^{9.23} {\rm GeV},\nonumber\\
\Delta_L(3,-1, 1)&\subset& {15}_H,\, M_{15_H}=10^{12}{\rm GeV},\nonumber\\
\xi(1,0,1)&&,~ M_{\xi} \sim {\cal O} (1) {\rm TeV}.\label{eq:kappadeltachi}
\eea
The scalar singlet $\xi(1,0,1)$ has played two 
crucial interesting roles of stabilising the SM scalar potential as
well as serving as WIMP DM candidate.
The introduction of ${15}_H$ at any scale $> 10^{9.23}$ GeV in this
model maintains precision coupling unification.
In the present  model we extend the model further by the inclusion of the
following fermions and an additional scalar $\chi_S(1,0,1)$     
\begin{itemize}
\item{Three right handed neutrino singlets $N_i(i=1,2,3)$,  one for
  each generation, with  masses to be fixed by this model phenomenology.}
\item{Three left-handed Majorana fermion singlets $S_i(i=1,2,3)$, 
one for each generation,  similar to those introduced in case of
inverse seesaw mechanism \cite{inv1,inv2,inv3,inv4,inv5,inv6,inv7}.}
\item{A Higgs scalar singlet $\chi_S(1,0,1)$  
  to generate $S-N$ mixings through its VEV.}
\end{itemize}
Being singlets under the SM gauge group,
they do not affect precision gauge coupling unification  of
ref.\cite{scp:2018}.
\subsection{Coupling Unification, GUT Scale, and Proton Lifetime}
As already discussed \cite{scp:2018,kynshi-mkp:1993} using renormalisation group equations for gauge couplings and the set of  Higgs scalars of eq(\ref{eq:kappadeltachi}),
precision unification has been achieved with the PDG values of input parameters
\cite{PDG:2012,PDG:2014,PDG:2016} on $\sin^2\theta_W(M_Z),\alpha_S(M_Z)$ resulting in the following 
mass scales and  the GUT
fine-structure constant $\alpha_G$ 
\bea
M_U&=& 10^{15.23} {\rm GeV}     , \nonumber\\
M_{\kappa}&=& 10^{9.23} {\rm GeV}, \nonumber\\
M_{\Delta_L}&=& M_{{15}_H}=10^{12} {\rm GeV},\nonumber\\
\frac{1}{\alpha_G}&=& 37.765. \label{eq:solsu5}
\eea

Using threshold effects due to superheavy Higgs scalars
\cite{Weinberg:1980,Hall:1981,Ovrut:1982,mkp:1987,mkp-cch:1989,rnm-mkp:1993,Langacker:1993,lmpr:1995},
 proton lifetime prediction for $p\to e^+\pi^0$ turns out to be in
the experimentally accessible range \cite{Abe:2017}
\be
\tau_p(p\to e^+\pi^0)= \left(1.01\times 10^{34\pm 0.44 }- (5.5\times
10^{35\pm 0.44 } \right) {\rm yrs}.\label{eq:taup}
\ee

Extensive investigations with number of SU(5) GUT extensions have been
carried out with
proton lifetime predictions
consistent with experimental limits
\cite{Dorsner,Dors-a,Dors-b,ppm:1989,Dors-c,Dors-d,Dors-e,Perez:2007,Nath-Perez:2007,Langacker:1981}. But implementation of type-II
seesaw dominance due to type-I seesaw cancellation resulting in
dominant LFV and LNV decays as dicussed below is new especially in the context
of non-SUSY SU(5).

\section{CANCELLATION OF TYPE-I AND DOMINANCE OF TYPE-II SEESAW }\label{sec:cancel}

Due to  introduction of heavy RH$\nu$s in the present model which were absent
in \cite{scp:2018}, it may be natural to presume
apriori that besides type-II seesaw, type-I
seesaw may also contribute substantially to light neutrino masses and mixings. But
it has been noted  that there is a natural mechanism to cancel out
type-I seesaw contribution while maintaining dominance of inverse
seesaw \cite{mkp-bs:2015,Kim-Kang:2006,app:2013,Majee:2009,mkp-ARC:2010,pas:2014} or type-II seesaw or even linear seesaw
\cite{nurev:mkpbpn,bpn-mkp:2015} as the case may be. Briefly we
discuss below how this cancellation mechanism operates in the present extended
model resulting in
  type-II seesaw dominance even in the presence of heavy RH$\nu$s.


The SM invariant Yukawa Lagrangian of the   model is 
\bea
{\cal L}_{\rm Yuk} &=& Y^{\ell} {\overline{\psi}}_{L}\psi_{R} \phi  
 +f \psi^c_{L} \psi_{L} \Delta_L \nonumber\\
&+& y_{\chi}{\overline N }^C S \chi_s 
+(1/2)M_{N}{\overline N}^C N + h.c. \label{eq:Yuk-Lag} 
\eea 
 Using the VEVs of the Higgs fields and denoting 
 $M=y_{\chi}\langle \chi_S \rangle =y_{\chi}V_{\chi}$,   
$M_D = Y\langle \phi \rangle$,
a $9 \times 9$ neutral-fermion mass matrix has been obtained which, upon
block diagonalization, yields  $3\times 3$ mass
matrices  for each of the light neutrino ($\nu_{\alpha}$), the right handed
 neutrino ($N_{\alpha}$), and the sterile neutrino
 ($S_{\alpha}$) \cite{app:2013,pas:2014,bpn-mkp:2015}.
The block diagonalisation of $9\times 9$ neutral fermion mass matrix
was  presented in  useful format in ref.\cite{LG:2000} but
without cancellation of type-I seesaw. Later on, this diagonalisation
procedure has been effectively utilized to study the type-I seesaw
cancellation mechanism in SO(10) models \cite{nurev:mkpbpn,app:2013,pas:2014,bpn-mkp:2015}.

In this model the left-handed triplet $\Delta_L$ and RH neutrinos
$M_N$ being much heavier than the other mass scales with 
$M_{\Delta_L}\gg M_N \gg M \gg M_D$ are at first integrated out from the Lagrangian leading to 
\begin{eqnarray}
- \mathcal{L}_{\rm eff} &=& \left(m_{\nu}^{II}+ M_D \frac{1}{M_N} M^T_D\right)_{\alpha \beta}\, \nu^T_\alpha \nu_\beta +
\left(M_D \frac{1}{M_N} M^T \right)_{\alpha m}\, \left(\overline{\nu_\alpha} S_m + \overline{S_m} \nu_\alpha \right)
\nonumber \\
&&\hspace*{4.0cm} +\left(M \frac{1}{M_N} M^T\right)_{m n}\, S^T_m S_n  \, ,\label{eq:effL}
\end{eqnarray}
which, in the $\left(\nu,~ S\right)$ basis, gives the $6 \times 6$ mass matrix
\begin{eqnarray}
\mathcal{M}_{\rm eff} = \left( \begin{array}{cc}
    M_DM_N^{-1} M^T_D+m_{\nu}^{II}  &  M_D M_N^{-1} M^T    \\
    MM_N^{-1} M_D^T       & MM_N^{-1} M^T  
        \end{array} \right) \, ,
\label{eqn:eff_numatrix}       
\end{eqnarray}
while the $3\times 3$ heavy RH neutrino mass matrix $M_N$ is the other part of the full 
$9 \times 9$ neutrino mass matrix. This $9 \times 9$ mass matrix $\tilde{\mathcal{M}}_{\rm 
\tiny BD}$ which results from the first step of block diagonalisation procedure as discussed 
above and in the appendix is
\begin{eqnarray}
\mathcal{W}^\dagger_1 \mathcal{M}_\nu \mathcal{W}^*_1 = 
\tilde{\mathcal{M}}_{\rm \tiny BD}
 = \bmt \mathcal{M}_{\rm eff} & 0\\
0& M_N
\emt \, .\label{eq:trMnu}
\end{eqnarray}
Defining
\bea
 X &=& M_D\,M^{-1},\nonumber\\
 Y&=&M\, M^{-1}_N,\nonumber\\
 Z&=&M_D\,M^{-1}_N,\label{eq:defXYZ}
\eea

the transfrmation matrix $\mathcal{W}_1$ has been derived as shown in
eqn.\,(\ref{app:w1})\cite{app:2013,bpn-mkp:2015}
\bea 
\mathcal{W}_1=\bmt
1-\frac{1}{2}ZZ^\dagger & -\frac{1}{2}ZY^\dagger & Z \\
-\frac{1}{2}YZ^\dagger & 1-\frac{1}{2}YY^\dagger & Y \\
-Z^\dagger & -Y^\dagger & 1-\frac{1}{2}(Z^\dagger Z + Y^\dagger Y)
\emt .
 \label{app:w1}
\eea

\noindent
After the second step of block 
diagonalization, the type-I seesaw contribution cancels out and gives in the $\left(\nu, S, N \right)$ basis
\begin{eqnarray}
\mathcal{W}^\dagger_2 \tilde{\mathcal{M}}_{\rm \tiny BD} \mathcal{W}^*_2 = \mathcal{M}_{\rm \tiny BD}
= \bmt m_\nu &0&0\\
0&m_{\cal S}&0\\
0&0&m_{\cal N}
\emt\, ,
\label{eqn:block-form}    
\end{eqnarray}
where $\mathcal{W}_2$ has been derived in eqn.\,(\ref{app:w2})
\cite{app:2013,bpn-mkp:2015}. We ahve used the bare mass of $S_i$ and
VEV of $\chi_L(2,-1/2,1)$ to be vanshing i,e $\mu_S=0,<\chi_L>=0$
to  get the form suitable for this model building
\bea 
\mathcal{W}_2 
=
\bmt 
1-\frac{1}{2}XX^\dagger &X & 0\\
-X^\dagger & 1-\frac{1}{2}X^\dagger X & 0 \\
0 & 0 & 1
\emt
 \label{app:w2}
\eea

In eq.(\ref{eqn:block-form}) the three $3 \times 3$ matrices are 
\bea
m_{\nu} & = & m_{\nu}^{II}=fv_L  \\ 
m_{\cal S}& = &  - M M^{-1}_NM^T  \\ 
m_{\cal N} &= &  M_N \, ,
\label{eq:mass}
\eea
the first of these being the well known type-II seesaw formula and the
second is the emergence of the corresponding type-I seesaw formula for
the singlet fermion mass. The third of the above equations represents the
heavy RH$\nu$ mass matrix.

In the third step, $m_{\nu}$, $m_{\cal S}$, and $m_{\cal N}$ are further diagonalised 
by the respective unitary matrices to give their corresponding eigenvalues
\begin{eqnarray}
U^\dagger_\nu\, m_{\nu}\, U^*_{\nu}  &=& \hat{m}_\nu = 
         \text{diag}\left(m_{1}, m_{2}, m_{3}\right)\, , \nonumber \\ 
U^\dagger_S\, m_{\cal S}\, U^*_{S}  &=& \hat{m}_S = 
         \text{diag}\left(m_{S_1}, m_{S_2}, m_{S_3}\right)\, , \nonumber \\
U^\dagger_N\, m_{\cal N}\, U^*_{N}  &=& \hat{m}_N = 
         \text{diag}\left(M_{N_1}, M_{N_2}, M_{N_3}\right)\, .
\label{eq:nudmass}
\end{eqnarray}
\noindent
The complete mixing matrix \cite{LG:2000,app:2013} diagonalising the above $9 \times 9$ 
neutrino mass matrix occuring  in 
eq.(\ref{eq:trMnu}) and in eq.(\ref{app:numass}) of Appendix  turns out to be
 \begin{eqnarray}
\mathcal{V}&\equiv&
\bmt 
{\cal V}^{\nu\hat{\nu}}_{\alpha i} & {\cal V}^{\nu{\hat{S}}}_{\alpha j} & {\cal V}^{\nu \hat{N}}_{\alpha k} \\
{\cal V}^{S\hat{\nu}}_{\beta i} & {\cal V}^{S\hat{S}}_{\beta j} & {\cal V}^{S\hat{N}}_{\beta k} \\
{\cal V}^{N\hat{\nu}}_{\gamma i} & {\cal V}^{N\hat{S}}_{\gamma j} & {\cal V}^{N\hat{N}}_{\gamma k} 
\emt \\
&=&\bmt 
\left(1-\frac{1}{2}XX^\dagger \right) U_\nu  & 
\left(X-\frac{1}{2}ZY^\dagger \right) U_{S} & 
Z\,U_{N}     \\
-X^\dagger\, U_\nu   &
\left(1-\frac{1}{2} \{X^\dagger X + YY^\dagger \}\right) U_{S} &
\left(Y-\frac{1}{2} X^\dagger Z\right) U_{N}   \\
0 &-Y^\dagger\, U_{S} & \left(1-\frac{1}{2}Y^\dagger Y\right)\, U_{N} 
\emt \, ,
 \label{eqn:Vmix-extended}
\end{eqnarray}
as shown in the Appendix. In eq.(\ref{eqn:Vmix-extended}) $X = M_D\,M^{-1}$, $Y=M\, 
M^{-1}_N$ and $Z=M_D\,M^{-1}_N$.

The mass of the singlet fermion is acquired through a type-I seesaw mechanism 

\bea
m_S=-M\frac{1}{M_N}M^T \label{matms}
\eea
where $M$ is the $N-S$ mixing mass term in the Yukawa Lagrangian  eq.(\ref{eq:effL}).

\section{ TYPE-II SEESAW FIT TO OSCILLATION DATA}\label{sec:t2fit}
\subsection{ Neutrino Mass Matrix from Oscillation Data}
Using diagonalisation of neutrino mass matrix $(m_\nu)$ by the PMNS matrix $U_{\rm PMNS}$
\begin{equation}
m_\nu = U_{\rm PMNS}~diag(m_1, m_2, m_3) U_{\rm PMNS}^T ,\label{mnu}
\end{equation}
where $m_i (i=1,2,3)$ denote the mass eigen values. For neutrino mixings we use the abbreviated cyclic notations $t_i=\sin\theta_{jk},c_i=\cos\theta_{jk}$ where $i,j,k$ are cyclic permutations of generational numbers $1,2,3$.
Following the standard parametrisation  we denote the PMNS matrix \cite{PDG:2012,PDG:2014,PDG:2016}
\begin{equation}
 U_{\rm PMNS}= \left( \begin{array}{ccc} c_{3} c_{2}&
                      t_{3} c_{2}&
                      t_{2} e^{-i\delta_D}\cr
-t_{3} c_{1}-c_{3} t_{1} t_{2} e^{i\delta_D}& c_{3} c_{1}-
t_{3} t_{1} t_{2} e^{i\delta_D}&
t_{1} c_{2}\cr
t_{3} t_{1} -c_{3} c_{1} s_{2} e^{i\delta_D}&
-c_{3} t_{1} -t_{3} c_{1} t_{2} e^{i\delta_D}&
c_{1} c_{2}\cr
\end{array}\right) 
{\rm diag}(e^{\frac{i \alpha_M}{2}},e^{\frac{i \beta_M}{2}},1),\label{eq:pmns}
\end{equation}
where  $\delta_D$ is the Dirac CP phase and $(\alpha_M,\beta_M)$ are Majorana phases. 

Here we present numerical analyses within $3\sigma$ 
limit of the neutrino oscillation data in the  type-II seesaw framework \cite{scp:2018}. As we do not have any experimental information 
about Majorana phases, they are determined by means of random
sampling: i,e  from the set of randomly generated values, each
confined within the maximum allowed limit of $2\pi$  only one set of
values for  $(\alpha_M,\beta_M)$ is chosen.  
Very recent analysis of the oscillation data has determined
the  $3\sigma$  and $1\sigma$ 
limits of Dirac CP phase $\delta_D$ \cite{Salas:2017} and there has
been 
 The best fit values 
of $\delta_D$ in the normally ordered (NO) and invertedly ordered (IO)
cases are near $1.2\pi$ and $1.5\pi$, respectively, which we utilise
for the sake of simplicity. A phenomenological model analysis has
yielded  $\delta_D=\pm 1.32\pi$ \cite{Ramond:2018}.

Global fit to the oscillation data \cite{Salas:2017}  is summarised below
 including respective parameter uncertainties at  $3\sigma$ level
\begin{eqnarray}
&&\theta_{12}\/^{\circ}=34.5\pm 3.25,\,\, \theta_{23}\/^{\circ}({\rm
    NO})=41.0\pm 7.25,  \nonumber \\
&& \theta_{23}\/^{\circ}({\rm IO})=50.5\pm 7.25,\,\theta_{13}\/^{\circ}({\rm NO})=8.44\pm0.5,\nonumber\\
&&\theta_{13}\/^{\circ}({\rm IO})=8.44\pm 0.5,\,\delta_{D}/\pi({\rm NO})=1.40\pm 1.0,\nonumber\\
 &&\delta_{D}/\pi({\rm IO})=1.44\pm 1.0,\nonumber \\
 &&\Delta m_{21}^2=(7.56\pm0.545)\times 10^{-5}{\rm eV}^2, \nonumber\\
&&|\Delta m_{31}|^2({\rm NO})=(2.55\pm 0.12)\times 10^{-3}{\rm eV}^2,\nonumber\\
&&|\Delta m_{31}|^2({\rm IO})=(2.49\pm 0.12)\times 10^{-3}{\rm eV}^2.
\label{oscdata}
\end{eqnarray}
We denote the cosmologically constrained parameter, the sum of the
three active neutrino masses, as
\be
\Sigma_{\nu}=\Sigma_i {\hat m}_i. \label{eq:sumnu}
\ee
For normally hierarchical (NH), inverted hierarchical (IH), and
quasi-degenerate (QD) patterns, the experimental values of mass
squared differences have been fitted by the following values of light
neutrino masses and the respective values of the cosmological
parameter $\Sigma_{\nu}$.
\bea
{\hat m}_{\nu}&=& (0.00127, 0.008838 ,0.04978) ~{\rm eV}\,\, (\rm
{NH})\nonumber\\   
\Sigma_{\nu}&=&0.059888~{\rm eV},\nonumber\\ 
{\hat m}_{\nu}&=& ( 0.04901,0.04978,0.00127)~{\rm eV}\,\, (\rm {IH})\nonumber\\
\Sigma_{\nu}&=&0.059888~~{\rm eV},\nonumber\\ 
{\hat m}_{\nu}&=& (  0.2056,0.2058,0.2) ~{\rm eV}\,\, (\rm {QD}),\nonumber\\
\Sigma_{\nu}&=&0.6114  ~~{\rm eV}. 
\label{eq:mnusumnu}
\eea
Using oscillation data and best fit values of the mixings we have also determined the PMNS mixing matrix
numerically
\bea
U_{\rm {PMNS}}=\begin{pmatrix} 0.816&0.56&-0.0199-0.0142i\\
-0.354-0.0495i&0.675-0.0346i&0.650\\
0.450-0.0568i&-0.485-0.0395i&0.75\end{pmatrix}.\label{eq:numupmns} 
\eea
\subsection{Determination of Majorana Yukawa Coupling Matrix} 

Now inverting the relation ${\hat m}_\nu=U_{PMNS}^\dagger {\mathcal M}_\nu U_{PMNS}^*$
where ${\hat m}_\nu$ is the diagonalised neutrino mass matrix, we determine ${\mathcal M}_{\nu}$ for three different cases and further determine the corresponding values of the $f$ matrix using $f={\mathcal M}_\nu/v_L$
where we use the predicted value of $v_L=0.1$ eV.\\ 
\par\noindent{\bf NH}\\
\bea
f =
\begin{pmatrix} 0.117+0.022i & -0.124-0.003i  &   0.144+0.025i\\
-0.124-0.003i  & 0.158-0.014i  & -0.141+0.017i\\ 0.144+0.025i &-0.141+0.017i& 0.313-0.00029i \end{pmatrix}\label{fNH}
\eea
\vspace{0.2cm}
\noindent{\bf IH}\\
\bea
f =  
\begin{pmatrix} 0.390-0.017i & 0.099+0.01i  &  -0.16+0.05i\\
0.099+0.01i  & 0.379+0.02i  & 0.176+0.036i\\-0.16+0.05i &0.176+0.036i& 0.21-0.011i \end{pmatrix}  \label{fIH}
\eea
\noindent{\bf QD}\\
\bea
 f =
\begin{pmatrix} 2.02+0.02i & 0.0011+0.02i  &  -0.019+0.3i\\
0.0011+0.02i  & 2.034+0.017i  & 0.021+0.21i\\-0.019+0.3i &0.021+0.21i& 1.99-0.04i \end{pmatrix} \label{fQD}
\eea

Randomly chosen Majorana phases \cite{scp:2018}
$\alpha_M=74.84^\circ,\beta_M=112.85^\circ$ and the central value of
the Dirac phase $\delta_D=218^\circ$ have been used in this analysis.
Using the well known definition of the Jarlskog-Greenberg \cite{Jarlskog:1985,Greenberg:1985} invariant 
\be
J_{CP}=-t_{3}c_{3}t_{2}c_{2}^2t_{1}c_{1} \sin \delta_D, \label{eq:jcp}
\ee
and keeping $\delta_D$ at its best fit values
we have estimated the predicted allowed ranges of the CP-violating parameter in both cases.
\bea
J_{CP}&=&0.0175-0.0212 \,\,(\rm NH)\, \nonumber\\
J_{CP}&=&0.0302-0.0365\,\,(\rm IH)\,\,\label{eq:numjcp}
\eea  
where the variables have been permitted to acquire values within their respective $3\sigma$ ranges of the oscillation data. Besides these there are non-unitarity contributions which have been discussed extensively in the literature.
\subsection{Scaling Transformation of Solutions}
In general there could be type-II seesaw models characterizing different seesaw scales and induced VEVs  matching the given set of neutrino oscillation data represented by the same neutrino mass matrix. For two such models
\bea
&&m_{\nu}=f^{(1)}v_L^{(1)} \nonumber\\
&&= f^{(2)}v_L^{(2)}, \label{eq:sc1}
\eea
Then the $f-$ matrix in one case is determined up to good approximation in terms of the other from the knowledge of the two seesaw scales\\
\be
f^{(1)} \simeq f^{(2)}\frac{M_{\Delta^{(1)}}}{M_{\Delta^{(2)}}}.\label{eq:sc2}
\ee
At $M_{\Delta^{(1)}}=10^{12}$ GeV our solutions are the same as in \cite{scp:2018}. 
In view of this scaling relation,  we can determine the values of the
Majorana Yukawa matrix in the present case from the estimations of
\cite{scp:2018}. For example, if we choose $M_{\Delta^{(1)}}=10^{10}$ GeV in the present case compared to
$M_{\Delta^{(2)}}=10^{12}$ GeV in \cite{scp:2018}, we rescale the solutions of
\cite{scp:2018} by a  factor $10^{-2}$ to derive solutions in the present case. 
Thus  graphical representations of solutions are similar to those of
ref.\cite{scp:2018} for $M_{\Delta^{(2)}}=10^{12}$ GeV which we do not
repeat here. The values of magnitudes of $f_{ij}$ at any new scale are
obtained by rescaling them by the appropriate scaling factor while the
phase angles remain the same as in \cite{scp:2018}.

\subsection{Dirac Neutrino Mass Matrix}
The Dirac neutrino mass matrix $M_D$ plays crucial role in predicting
LFV and LNV decays. In certain SO(10) models
\cite{babu-rnm:1993,Joshipura,bpn-mkp:2015,ap:2012} this is usually determined by
fitting the charged fermion masses at the GUT scale and equating it with the upquark mass matrix. The fact that $M_D^0\simeq M_{u}^0$ at the GUT scale follows from the underlying quark lepton symmetry \cite{JCP:1974} of SO(10). In SU(5) itself, however, there is no
such symmetry to predict the structure of $M_D$ in terms of quark matrices. Also
in this SU(5) model we do not attempt any charged fermion mass fit at the GUT
scale or above it. Since the Dirac neutrino mass matrix is not predicted by the SU(5) symmetry itself, for the sake of simplicity and to derive maximal effects on LFV and LNV decays, we  assume $M^0_D$ to be equal to the up-quark mass
matrix $M_u^0$ at the GUT
scale. Noting that $N$ is SU(5) singlet fermion, in the context of
relevant Yukawa interaction Lagrangian 
\be
-{\cal L}_{\rm Yuk}= [Y_{N}{\overline 5}_F.{\bf 1}_F.{5}_H + Y_{u} {10}_F.{10}_F.{5}_H+....]+h.c.,\label{eq:su5yuk}
\ee
this assumption is equivalent to alighment of the two Yukawa couplings
\be
Y_{N}\simeq Y_{u} . \label{eq:eqyuk}
\ee 
This alignment is naturally predicted in SO(10) or
SO(18)\cite{Wilczek-Valle}, but in the present SU(5) case it is assumed.

We realise this matrix $M_D$ using renormalisation group equations for fermion masses and gauge couplings and their numerical solutions \cite{dp:2001,dp-a,dp-b} starting from the PDG values \cite{PDG:2012,PDG:2014,PDG:2016} of fermion masses at the electroweak scale. 
Following the bottom-up approach and using the down quark diagonal basis, the quark masses and the CKM mixings are 
extrapolated from low energies using renormalisation group (RG) equations
\cite{dp:2001,dp-a,dp-b,Stella:2016}. After assuming the approximate equality $M^0_D\simeq M^0_u$ at the GUT scale where $M^0_u$ is the up-quark mass matrix,  the top-down approach is
exploited to run down this mass matrix $M^0_D$ using  RG equations
\cite{dp:2001}. Then
   the value of $M_D$ near $1-10$ TeV scale turns out to be  
\bea
M_D \simeq\footnotesize\begin{pmatrix}
0.014&0.04-0.01i&0.109-0.3i\\
0.04+0.01i&0.35&2.6+0.0007i\\
0.1+0.3i&2.6-0.0007i&79.20
\end{pmatrix}{GeV}.  \label{eq:mdmu}
\eea

As already noted above, although on the basis of SU(5) symmetry alone there may not be any reason for the rigorous validity of eq.(\ref{eq:mdmu}), in what follows  we study the implications of this assumed value of $M_D$ to examine maximum possible impact on LFV and LNV decays discussed in Sec.\ref{sec:lfv} and Sec.\ref{sec:bb}. Another
reason is that the present assumption on $M_D$ may be justified in direct SO(10)
breaking to the SM which we plan to pursue in a future work.

\section{RIGHT-HANDED NEUTRINO  MASS IN SU(5) vs. SO(10)}\label{sec:rhnu}
\subsection{RH$\nu$ Mass in SO(10)}
The fermions responsible for type-I and type-II  seesaw are the  LH
leptonic doublets and the RH fermionic singlets of three generations.
 In SO(10) the left handed lepton doublet  $(\nu, l)^T$
and the right-handed neutrino $N$ are in the same spinorial representation ${16}_F$.
\bea
(\nu, l)^T & \subset & {16}_F,\nonumber\\
N& \subset & {16}_F.\label{eq:fermiso10}
\eea
The Higgs representation ${126}^{\dagger}_H\subset SO(10)$ contains  both the left-handed and the right-handed triplets carrying $B-L=-2$,
\be
{126}^{\dagger}_H=\Delta_L(3,1,-2,1)+\Delta_R(1,3,-2,1)+.....\label{eq:126h}\\
\ee
where the quantum numbers are under the left-right symmetry group $SU(2)_L\times SU(2)_R \times U(1)_{B-L}\times SU(3)_C$. 
The common Yukawa coupling $f_{10}$ in the Yukawa term
\be
-{\cal L}_{\rm Yuk10}=  f_{10} {16}_F{16}_F{126}_H^{\dagger},\label{eq:Yuk10}
\ee
 generates the dilepton-Higgs triplet interactions both in the left-handed and right-handed sectors giving rise to type-I and type-II seesaw mechanisms.
The RH neutrino mass is generated through the VEV of the neutral component of the of $\Delta_R$
\be
M_N=f_{10} \langle \Delta_R^0 \rangle .\label{eq:MNso10}.
\ee
The type-II seesaw contribution to light  neutrino mass is 
\be
{\cal M}_{\nu}=f_{10} v_L  \label{eq:t2so10}
\ee
where $v_L$ is the corresponding induced VEV of $\Delta_L$
\be
v_L=\lambda_{10}\frac{\langle \Delta_R^0\rangle v^2_{ew}}{M_{\Delta_L}^2}. \label{eq:vlso10}
\ee
Here $\lambda_{10}$ is the quartic coupling in the part of the scalar
potential
\bea
 V_{10}&=&\lambda_{10}\Delta_L^{\dagger}\Delta_R \phi^{\dagger} \phi \nonumber\\
&\subset& \lambda_{10}{126}_H^{\dagger}{126}_{H}{10}_H{10}_H.\label{eq:v10}
\eea 
Thus with type-II seesaw dominance, the predicted heavy RH neutrino masses in SO(10)  follow the same hierarchical pattern as the active light neutrino masses
\be
M_{N_1}:M_{N_2}:M_{N_3}::m_1:m_2:m_3 .\label{eq:rhmso10}
\ee
\subsection{RH$\nu$ Mass in SU(5)}

\begin{figure}[htbp]
 \includegraphics[width=6cm,height=4.5cm,angle=0]{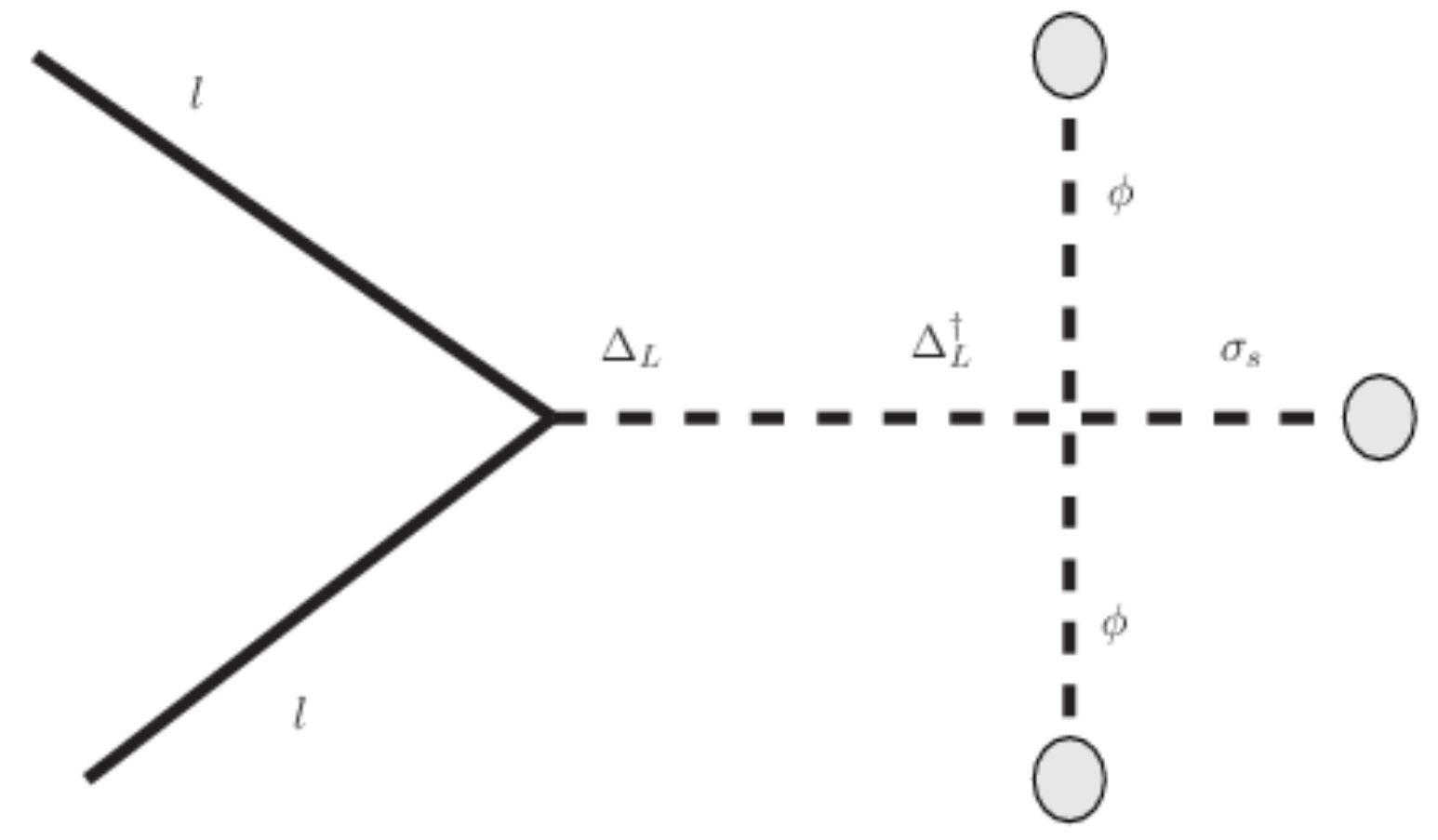}
 \caption{Feynman diagram representing type-II seesaw mechanim for
   neutrino mass generation in SU(5). Scalar fields $\phi, \sigma_S$ and
   $\Delta_L$ represent SM Higgs doublet, singlet, and LH triplet as
   defined in the text. This diagram defines the trilinear coupling
   mass $\mu_{\Delta}=\lambda \langle \sigma_S \rangle$.}   
\label{fig:su5mod2}
\end{figure}

Feynman diagram for type-II seesaw mechanism in the present SU(5) model is
shown in Fig.\ref{fig:su5mod2}.
In contrast to SO(10) where the LH leptonic doublet and the RH$\nu$ are
in one and the same representation ${16}_F$, in SU(5) they are in different 
representations
\bea
(\nu, l)^T & \subset & {\overline 5}_F,\nonumber\\
N& \subset & {\bf 1}_F.\label{eq:fermiso10}
\eea
In SU(5), while the dilepton Higgs interaction is given by
\be
-{\cal L}_{Yukll}=f {\bar 5}_F{\bar 5}_F{15}_H,\label{eq:Yukll5}
\ee
the RH neutrino mass is  generated through
\be 
-{\cal L}_{YukNN}= (1/2)f_NN N \sigma_S +h.c.\label{eq:YukNN5}
\ee 
The fact that $N$ is a singlet under SU(5) forces $\sigma_S$ to be a singlet too . Further 
this singlet $\sigma_S$ must carry $B-L=-2$ as its VEV generates the
heavy  Majorana mass
\be 
M_N=f_N\langle \sigma_S \rangle .\label{eq:MN}
\ee
In sharp contrast to SO(10) where the LH triplet $\Delta_L$
and the RH triplet $\Delta_R$ scalars contained in the same representation
${126}_H^{\dagger}$ generate the type-II seesaw and $M_N$, the
situation in SU(5) is different. Since  LH
triplet $\Delta_L(3,-1,1)$ mediating type-II seesaw belongs to Higgs
representation ${15}_H\subset $ SU(5) and $\sigma_S$ belongs to a completely
different representation (which is a singlet ${\bf 1}\subset$ SU(5)), the
two relevant Majorana type couplings in general may not be equal
\be
f_N\neq f.               \label{eq:unequal}
\ee
Also this assertion is further strengthened if we do not assume SU(5)
to be a remnant of SO(10).
Then the RH neutrino mass hierarchy can be decoupled from  the type-II seesaw prediction.
It is interesting to note that in SU(5)
\be
v_L=\frac{\mu_{\Delta}v_{\rm ew}^2}{M_{\Delta}^2}
\ee
where $\mu_{\Delta}$ is the trilinear coupling in the potential term
\be
V_{tri}=\mu_{\Delta}\Delta_L \phi \phi +h.c. \label{eq:tlpot}
\ee
 The VEV of this singlet $\sigma_S$ can explain the dynamical origin of such trilinear coupling through its VEV $v_{\sigma}=\langle \sigma_S \rangle$ 
\be
\mu_{\Delta}=\lambda v_{\sigma}, \label{eq:model}
\ee
where $\lambda$ is the quartic coupling in the potential term
\bea
V_{ql}&=&\lambda \sigma_S\Delta_L\phi\phi + h.c.\\
&\subset&\lambda \sigma_S {15}_H{5}_H{5}_H+h.c \label{eq:qpot}
\eea
where the second line represents the SU(5) origin. For GUT-scale
$U(1)_{B-L}$ symmetry breaking driving VEV $v_{\sigma}\simeq M_{GUT}$
in the natural absence of any intermediate symmetry, it is possible to ensure $\mu_{\Delta}\simeq M_{\Delta_L}$ for 
\be
\lambda \simeq \frac{M_{\Delta_L}}{M_{GUT}}. \label{eq:lamdelgut}
\ee
Thus the SU(5) model gives similar explanation for quartic coupling as
in direct breaking case of SO(10).
But the predicted hierarchy of RH$\nu$ masses in SU(5) may not, in general,
follow the same hierarchical pattern as given by SO(10)  shown in eq.(\ref{eq:rhmso10}).
This is precisely so because  eq.(\ref{eq:rhmso10}) follows from the
fact that the
same  matrix $U_{PMNS}$ diagonalises both the LH and the RH
neutrino mass matrices which is further rooted in the fact that same
Majorana coupling $f_{10}$ that generates the type-II seesaw mass term
also generates $M_N$. But because of the general possibility in SU(5) that $f_N\neq
f$, the RH$\nu$s may acquire a completely different pattern depending
upon the value of $f_N$. Unlike
SO(10), these masses emerging from SU(5) are also allowed to be quite
different from the type-II seesaw scale.

Even if the value of $v_{\sigma}$
may be needed to be near $M_{\Delta_L}$, the value of $M_N$ is allowed
to be considerably lower by finetuning the value of $f_N$.
Our LFV and LNV decay phenomenology as discussed below may need $M_N=1-10$ TeV which is realizable using this new technique in SU(5). In contrast SO(10) needs
$U(1)_R\times U(1)_{B-L}$\cite{ap:2012,mkp-bs:2015,bpn-mkp:2015} or $SU(2)_R\times U(1)_{B-L}$ gauge symmetry
and hence new gauge bosons  near the TeV scale to generate such RH$\nu$
masses which should be detected at LHC \cite{mkp-bs:2015,bpn-mkp:2015}.   
Thus a new mechanism for RH$\nu$ mass emerges here by noting the
coupling $f_N\neq f$ which has the potential to generate RH$\nu$
masses over a wide range of values $M_N\sim 100-10^{15}$ GeV. Then the RH$\nu$ mass
predictions in the two GUTs in the presence of type-II seesaw
dominance are
\par\noindent{\bf{\underline {Type-II Seesaw Dominated SO(10)}:-}}\\
\be
M_{N_i}\simeq \frac{m_i M^2_{\Delta_L}}{ v_{ew}^2}. \label{eq:mnso10}
\ee
\par\noindent{\bf{\underline{Type-II Seesaw Dominated SU(5)}:-}}\\
\be
M_{N_i}=\left[{\cal O}(10){\rm GeV} - {\cal
    O}(M_{\Delta_L})\right].\label{eq:mnsu5}
\ee 
Here $m_i,i=1,2,3$ are the three mass eigen values of light
neutrinos. It is to be noted that $m_i$ is absent in the RHS of
eq.(\ref{eq:mnsu5}) in the SU(5) case.
\subsection{Realization of Mass Hierarchies}\label{sec:hier}
Here we discuss how the stated hierarchy in Sec.\ref{sec:cancel}  
\be
M_{\Delta_L}\gg M_N \gg M \gg M_D, \label{eq:t2hier}
\ee
for type-II seesaw dominance is realized through fine tuning.
At first noting that the mass squared term for ${15}_H\subset SU(5)$ in
the scalar potential, $M_{15}^2{15}_H^{\dagger}{15}_H$, is SU(5)
invariant, $M_{15}$ can have any mass below the GUT scale subject to
proton stability and gauge coupling unification. Since ${15}_H$,
unlike ${5}_H$, does not have Yukawa interactions with SM fermions, the
Higgs mediated proton decay is suppressed. We now explain why we have
used $M_{15}=M_{\Delta_L}=10^{12}$ GeV. In our model it is possible
to assign any mass $M_{15}=M_{\Delta_L} >M_{\eta}$ where the mass of
$\eta(3,0,8)\subset {75}_H$ is
$M_{\eta}=10^{9.23}$ GeV. Because of the presence of $\eta(3,0,8)$ at
such intermediate mass value, precision gauge coupling unification is achieved which
has been discussed separately
\cite{kynshi-mkp:1993,scp:2018}. Following the standard symmetry
breaking SU(5)$\to$ SM through the GUT scale VEV of the SM singlet
scalar in the adjoint representation ${24}_H$,
$V_{GUT}=\langle{24}_H^0\rangle>$, 
 a SU(5) invariant scalar potential $V_{\eta}$ gives the mass of $\eta(3,0,8)$ 
\bea
V_{\eta}&=&M_{75}^2{75}_H^2+m_{(24,75)}{24}_H{75}_H^2
+\lambda_{(24,75)}{24}_H^2{75}_H^2\nonumber\\
&\supset&\left[M_{75}^2+m_{(24,75)}V_{GUT}+\lambda_{(24,75)}V_{GUT}^2\right]\eta^2.\label{eq:etamass}
\eea
leading to
$M_{\eta}^2=M_{75}^2+m_{(24,75)}V_{GUT}+\lambda_{(24,75)}V_{GUT}^2$. 
Here $M_{75}\sim m_{(24,75)} \sim
V_{GUT}\sim M_{GUT}$. Thus fine tuning any
one of these four parameters can give $M_{\eta}=10^{9.23}$ GeV.
The presence of ${15}_H $ below $M_{\eta}$ destabilises unification but
protects it for $M_{15} > M_{\eta}$. This has led to the chosen value
of $M_{\Delta_L}=M_{15}=10^{12}$ GeV. We have noted in the following
section that the value of
$M_3 < 1$ TeV violates the observed bound on the non-unitarity
parameter $\eta_{\tau\tau}< 2.7\times 10^{-3}$ leading to the lower
bound on $M$ in the degenerate case
\be
M_1=M_2=M_3\ge 1250\,\, {\rm GeV}. \label{eq:Mbound}
\ee
where $M={\rm diag} (M_1,M_2,M_3)$. 
Noting the definition 
\be
M=y_{\chi}\langle \chi_S(1,0,1) \rangle
=y_{\chi}V_{\chi},\,\,\label{eq:defM}
\ee
 we now argue that even for GUT scale  mass of
$\chi_S$ and its VEV $V_{\chi}=V_{GUT}$,
 it is possible to satisfy eq.(\ref{eq:Mbound}). For
$V_{\chi}\sim V_{GUT}\sim 10^{15}$ GeV we need a small fine tuned value of
Yukawa coupling
\be
y_{\chi} > 10^{-12}, \label{eq:yukchi}  
\ee
which satisfies $M \ge 1$ TeV but does not affect any known fermion mass. This shows the
interesting possibility that even without having a low mass
non-standard Higgs $\chi_S$, it is possible to realize the extended
seesaw with type-II seesaw dominance.
However, if we insist on $y_{\chi}\le 1$, we need $V_{\chi}\ge 1$
TeV and $M_{\chi_S} \ge 1$ TeV which is realisable as $\chi_S$ is a SU(5)
scalar singlet. 
As we have assumed $M_D=M_u$, this gives at GUT scale on extrapolation\cite{dp:2001}  
\be
M_{D_{33}}\sim m_{\rm  top}\simeq 85\,\, {\rm GeV}.\label{eq:MD33}
\ee
where $m_{\rm top}$ is the top-qurk mass.
Thus achieving precision unification and type-II seesaw dominance has
already given us
$M_{Delta}>M_{\eta}$ whereas fine tuning the Majorana coupling, $f_N >
10^{-11}$, has yielded $M_N > 10^4$ GeV. Combining these with
eq.(\ref{eq:Mbound}) and eq.(\ref{eq:MD33}) gives the hierarchical
relation of eq.(\ref{eq:t2hier}).
 
\section{\bf LEPTON FLAVOR VIOLATIONS}\label{sec:lfv}
Using SM extensions there has been extensive investigation of lepton
flavor violating phenemena $l_{\alpha} \to l_{\beta}+\gamma,
(\alpha\neq \beta) $ and other
processes like $\mu \to e{\bar e}e$  including
unitarity violations \cite{ilakovac,Depp:2005,Arina:2008,Mal:2009,Hirsch:2009,Depp:2006,spdas:2012,antbnd,antu-a,antu-b,antu-c,non-unit,non-unit-a,non-unit-b,non-unit-c,non-unit-d,non-unit-e}. In the flavor
basis we use the
standard charged current Lagrangian  
\begin{eqnarray}
\mathcal{L}_{\rm CC} &=& -\frac{1}{\sqrt{2}}\, \sum_{\alpha=e, \mu, \tau}
[g_{2L} \overline{\ell}_{\alpha \,L}\, \gamma_\mu {\nu}_{\alpha \,L}\, W^{\mu}_L] 
       + \text{h.c.} 
\label{eqn:ccint-flavor}
\end{eqnarray}

In  predicting the LFV
branching ratios  we have used the relevant formulas of
\cite{ilakovac} and  assumed a simplifying diagonal structure for
M,
\be
M = {\rm diag.}~(M_1, M_2, M_3).\label{eq:M}
\ee
Then eq.(\ref{eq:M}) in combination with eq.{\ref{eq:mdmu}), gives the elements of the
  $\nu-S$ mixing matrix 
\bea
{\cal V}^{(lS)}=
\begin{pmatrix}
{M_D}_{e1}/M_1& {M_D}_{e2}/M_2&{M_D}_{e3}/M_3\\ 
{M_D}_{\mu 1}/M_1& {M_D}_{\mu 2}/M_2&{M_D}_{\mu 3}/M_3\\ 
{M_D}_{\tau 1}/M_1& {M_D}_{\tau 2}/M_2&{M_D}_{\tau 3}/M_3
\end{pmatrix}.
\label{HLMIX}
\eea
The $S-N$ mixing matrix 
\be
V^{(SN)}=\frac{M}{M_N}, \label{eq:mixsn} 
\ee
is relatively damped out since $M_N\gg M$. In fact the type-I
cancellation condition $M_N >> M >> M_D$ ensures this damping. 
Noting that the physical neutrino flavor state $\nu_{\alpha}$ is a mixture of ${\hat \nu}$, ${\hat S}$ and ${\hat N}$
\be
 \nu_{\alpha}=U_{\alpha i}{\hat \nu}_i+V^{lS}_{\alpha i}{\hat S}
+V^{(SN)}_{\alpha i}{\hat N_i}.\label{eq:nusnmix}
\ee
where $U\sim U_{PMNS}$ and the other two mixings violate unitarity. For large $M_N\gg M$ the third term in the RHS of eq.(\ref{eq:nusnmix}) can be dropped leading to the unitarity violation parameter $\eta$
\be
U^{\prime}\simeq(1-\eta)U_{PMNS}.\label{eq:nonuni}
\ee
where
\bea
\eta_{\alpha\beta}&=&(1/2){(X.X^{\dagger})}_{\alpha\beta},\nonumber\\
X&=&\frac{M_D}{M}. \label{eq:X}
\eea
There has been extensive discussion on the constraint imposed on this
parameter
\cite{antbnd,non-unit}. The largest out of these is $\eta_{\tau \tau}\le 0.0027$. Theoretically
\be
\frac{1}{2}\left[\sum_i \frac{{M_D}_{\tau i}.{M_D}_{\tau i}^*}{M_i^2}\right]\le 0.0027.\label{eq:etanum}
\ee
In the completely degenerate case of $S-N$ mixing, $M_1=M_2=M_3=M$ we get
\be  
M\ge 1250 {\rm GeV} \label{eq:Mdeg}
\ee

 The  RH neutrinos in the present  model being degenerate with masses
 $M_{N_i}\gg m_{S_i}$ have much less significant contributions than the singlet fermions. The predicted branching ratios 
being only few to four orders less than the current experimental
limits \cite{lfvexpt} are verifiable by ongoing searches,

\bea
BR(\mu \to e\gamma)&=&1.19\times 10^{-16},\nonumber\\
BR(\tau \to e\gamma)&=&1.69\times 10^{-14},\nonumber\\
BR(\tau \to \mu\gamma)&=&1.8\times 10^{-12}.\label{lfvbr}
\eea 
For the sake of completeness we present the variation of LFV decay branching ratios as a function of the lightest neutrino mass in Fig. \ref{lfvbrs}. \\     

  \begin{figure}[htbp]
 \includegraphics[width=10cm,height=8cm,angle=0]{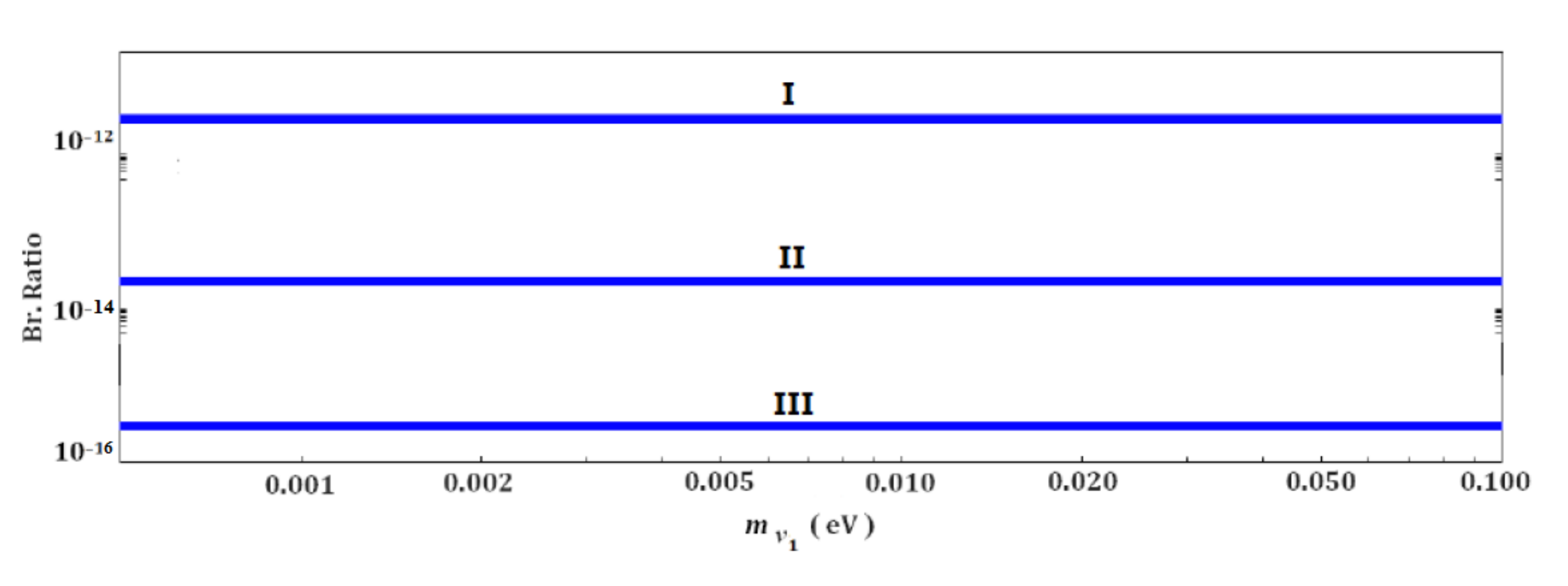}
 \caption{Variation of LFV decay branching ratios as a function of the
   lightest neutrino mass. Colored horizontal lines represent ${\rm I}:BR(\tau \to
   \mu\gamma)$, ${\rm II}:BR(\tau \to e\gamma)$, and ${\rm III}:BR(\mu \to e\gamma)$. }   
\label{lfvbrs}
\end{figure}

 In this approach the LFV decay rate
mediated by the $W_L$ boson in the loop depends predominantly upon $N-S$ mixing matrix $M$
and the Dirac neutrino mass matrix $M_D$, although subdominantly upon
the RH$\nu$ mass matrix $M_N$. However in the high scale type-II seesaw
 ansatz followed here
 LFV decay rate is independent of  light neutrino masses. This behavior
 of LFV decay rates are clearly exhibited in Fig.\ref{lfvbrs} where
 the three branching ratios have maintained constancy with the
 variation of $m_{\nu}$.


\section{\bf DOMINANT $W_L-W_L$-CHANNEL DOUBLE BETA DECAY WITHIN COSMOLOGICAL BOUND}\label{sec:bb}
\subsection{Double Beta Decay Mediation by Sterle Neutrinos}  

In the absence of $W_R$ bosons and right-handed $\Delta_R^{\pm\pm}$ in
 SU(5), there is no contribution to right-handed double beta
 production. The gauge coupling unification constraint sets the
lower bound on the masses of  left-handed doubly charged Higgs bosons
$\Delta_L^{\pm\pm}$ to be $M_{\Delta_L}\simeq M_{{15}_H} > 10^{9.23}$ GeV. 
As such they have negligible contributions for direct mediations of
$0\nu\beta\beta$ process with left-handed electrons. Thus the only significant
contributions in the $W_L-W_L$ channel could be through the mediation
of  $\nu$, $S$, and $N$. 
 Feynman diagrams for
$0\nu\beta\beta$ decay amplitude in the $W_L-W_L$ channel due to the exchanges of Majorana
fermions $\nu$, $S$, and $N$ are shown in Fig. \ref{fig:feynall}.
\begin{figure}[htbp]
 \includegraphics[width=15cm,height=5cm,angle=0]{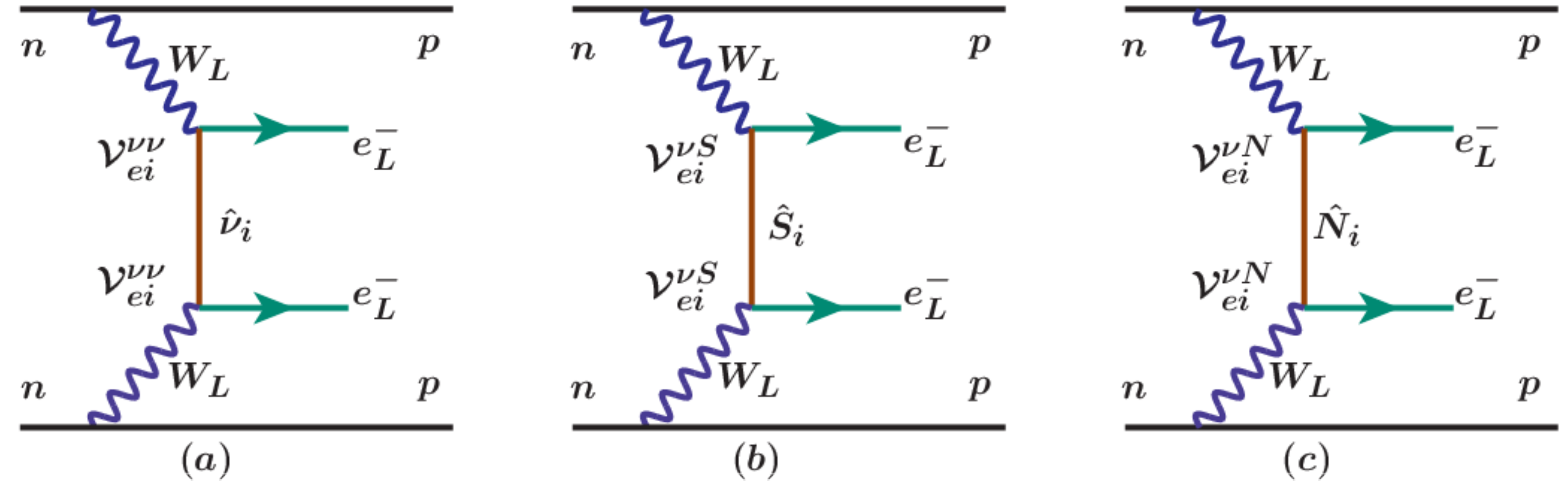}
 \caption{Feynman diagrams representing neutrino-less double beta
   decay due to
exchanges of all three types of  Majorana frmions $\nu$,$S$ and $N$.}   
\label{fig:feynall}
\end{figure}
In Fig.\ref{fig:feyns} we also present  Feynman diagram for
$0\nu\beta\beta$ decay amplitude due to the sterile neutrino  exchange 
where its mass insertion has been explicitly indicated.
\begin{figure}[htbp]
 \includegraphics[width=6cm,height=4.5cm,angle=0]{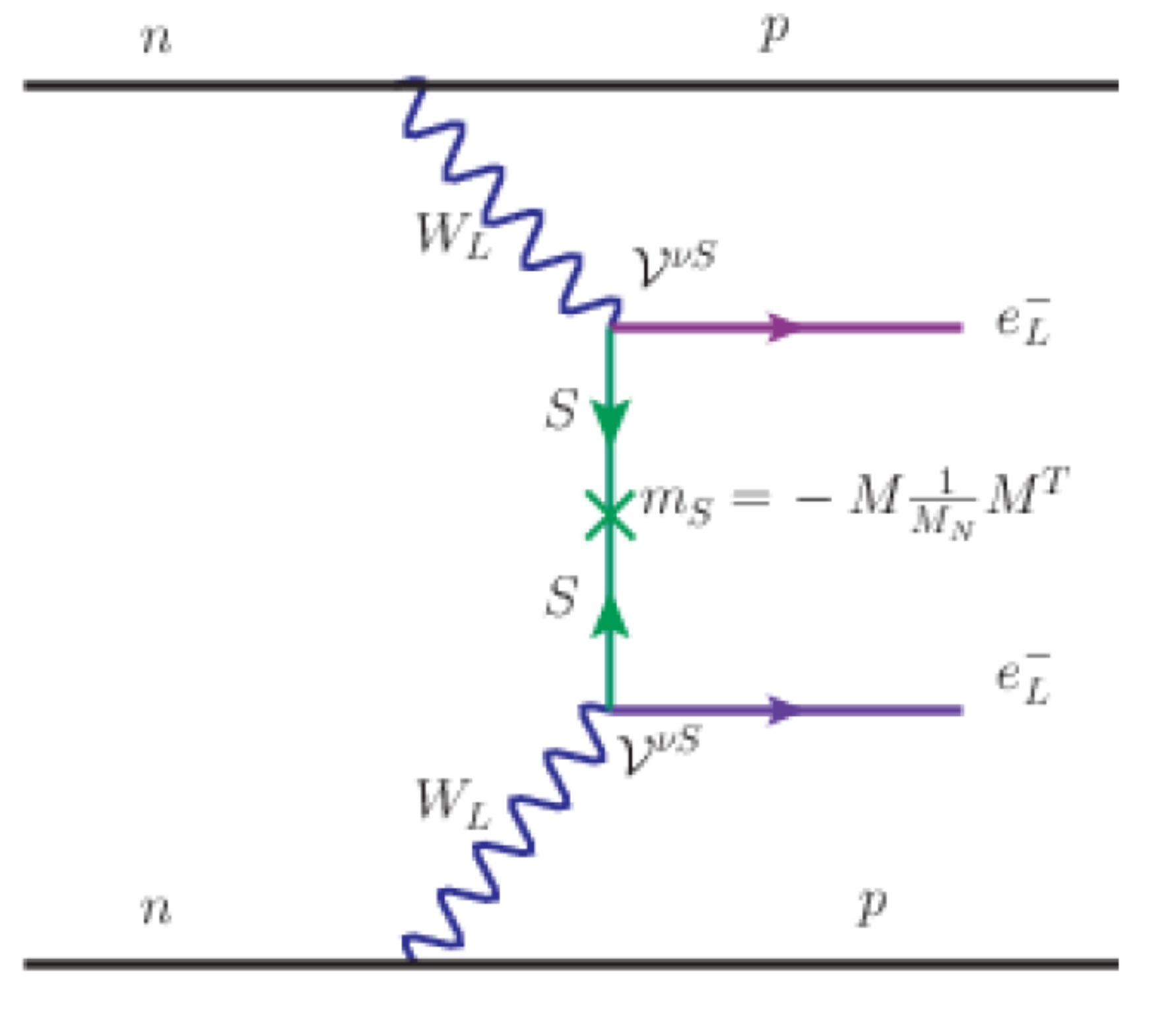}
 \caption{Feynman diagram representing neutrino-less double beta
   decay amplitude due to
exchanges of singlet fermions $S_i(i=1-3)$ with explicit mass isertion $m_s=-M\frac{1}{M_N}M^T$.}   
\label{fig:feyns}
\end{figure}
Mass eigen values of different sterile neutrinos for different sets of 
 $(M_1,M_2,M_3)$ consistent  with constraints on unitarity violating parameters 
$\eta_{\alpha \beta}$ are presented in Table \ref{tab:msMi}. We have
used the singlet fermion  mass seesaw formula of eq.(\ref{eq:mass})
and $M_{N_1}=M_{N_2}=M_{N_3}= 4000$ GeV. 
\begin{table}[!h]
\caption{Prediction of singlet fermion masses for different values of
  $(M_1,M_2,M_3)$ where we have used  $M_{N_1}=M_{N_2}=M_{N_3}= 4000$ GeV. }
\begin{center}
\begin{tabular}{|c|c|}
\hline
{ $M$ (GeV)} & {$\hat{m}_s$ (GeV)} \\
\hline
$(60,1200,1200)$ & $(0.9,360,360)$ \\
\hline
$(70,1200,1200)$ & $(1.22,360,360)$ \\
\hline
$(80,1200,1200)$ & $(1.60,360,360)$ \\
\hline
$(90,1200,1200)$ & $(2.00,360,360)$ \\
\hline
$(100,1200,1200)$ & $(2.50,360,360)$ \\
\hline
$(110,1200,1200)$ & $(3.00,360,360)$ \\
\hline
$(120,1200,1200)$ & $(3.60,360,360)$ \\
\hline
$(130,1200,1200)$ & $(4.22,360,360)$ \\
\hline
$(140,1200,1200)$ & $(4.90,360,360)$ \\
\hline
$(150,1200,1200)$ & $(5.62,360,360)$ \\
\hline
\end{tabular}
\end{center}
\label{tab:msMi}
\end{table}
These solutions are displayed in Fig.\ref{fig:msM}. 
\begin{figure}[htbp]
 \includegraphics[width=9cm,height=12cm,angle=-90]{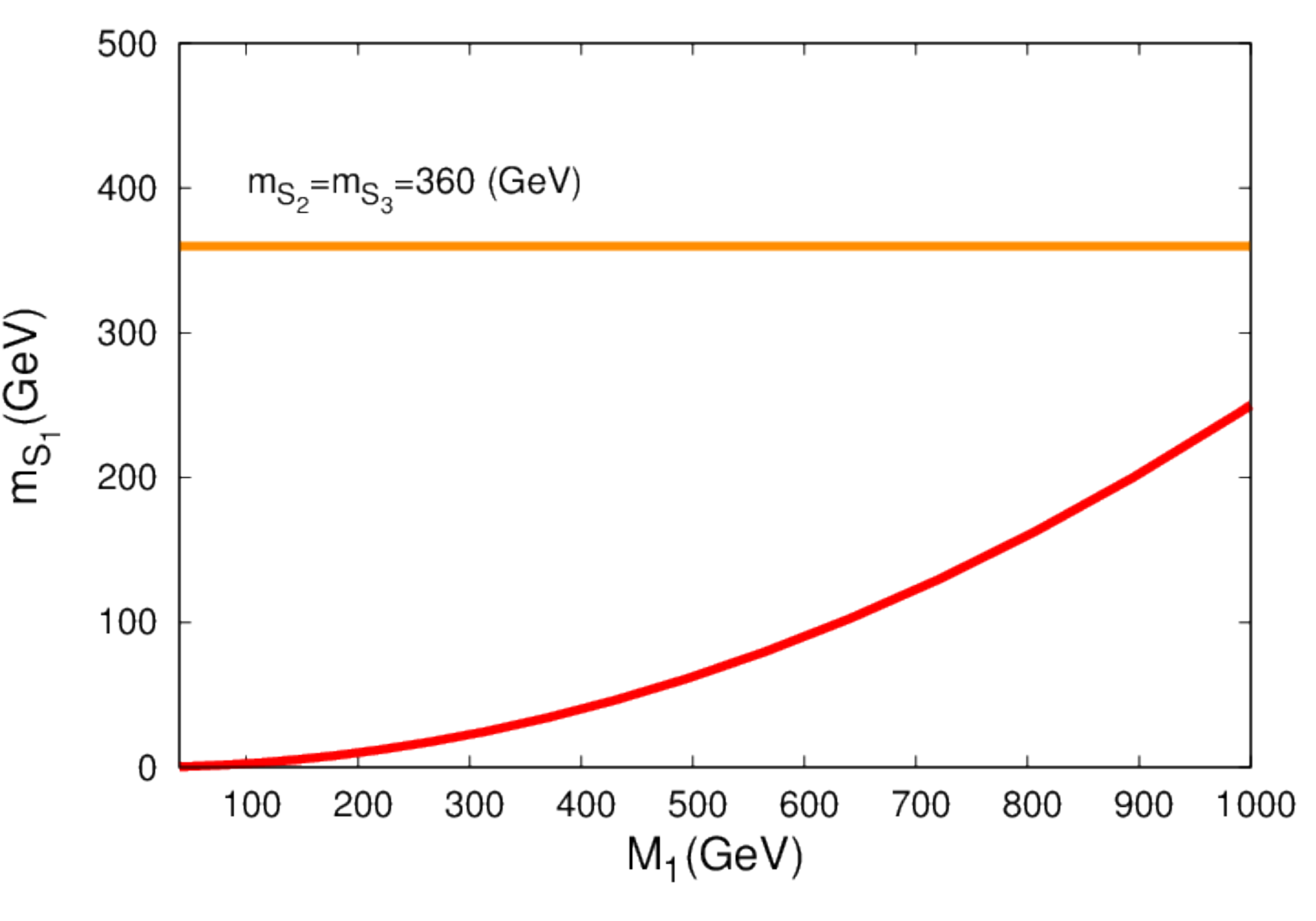}
 \caption{Prediction of singlet fermion mass eigen values as a function of $N-S$ mixing mass parameters $M_i(i=1,2,3)$ for $M_{N_i}= 4\, {\rm TeV} (i=1,2,3)$. The horizontal red coloured line represents solutions for two other eigen values for $M_2=M_3=1200$ GeV.} 
 \label{fig:msM}
\end{figure}

We use normalisations necessary for different contributions \cite{Vergados,Ver-a,Ver-b,Doi,Doi-a,Doi-b,Doi-c} due to exchanges of
 light-neutrinos, sterile neutrinos, and the heavy
RH neutrinos  in the $W_L-W_L$ channel. They 
lead to the inverse half life \cite{app:2013,pas:2014,bpn-mkp:2015},   
\begin{eqnarray}
  \left[T_{1/2}^{0\nu}\right]^{-1} &\simeq &
  G_{01}|\frac{{\cal M}^{0\nu}_\nu}{m_e}|^2|({\large\bf
    M}^{ee}_{\nu} +{\large\bf M}^{ee}_{S}+{\large\bf M}^{ee}_{N})|^2,\nonumber\\
&=& K_{0\nu}|({\large\bf
    M}^{ee}_{\nu} +{\large\bf M}^{ee}_{S}+{\large\bf M}^{ee}_{N})|^2,\nonumber\\
&=& K_{0\nu}|{\large\bf
    M}_{\rm eff} |^2.
  \label{invhalf}
\end{eqnarray}
Here  $G_{01}= 0.686\times 10^{-14} {\rm yrs}^{-1}$, ${\cal
  M}^{0\nu}_{\nu} = 2.58-6.64$, and $K_{0\nu}= 1.57\times 10^{-25} {\rm yrs}^{-1}
{\rm eV}^{-2}$. In eq.(\ref{invhalf}) the three effective mass parametes  have been defined as 
\begin{eqnarray}
{\large \bf  M}^{\rm ee}_{\nu} =\sum^{}_{i} \left(\mathcal{V}^{\nu \nu}_{e\,i}\right)^2\, {m_{\nu_i}}
\label{effmassparanus} 
\end{eqnarray}
\bea
{\large \bf  M}^{\rm ee}_{S} = \sum^{}_{i} \left(\mathcal{V}^{\nu
  S}_{e\,i}\right)^2\, \frac{|p|^2}{{\hat m}_{S_i}} 
  \label{effmassparanus2} 
\eea
\begin{eqnarray}
{\large \bf  M}^{\rm ee}_{N} = \sum^{}_{i} \left(\mathcal{V}^{\nu
  N}_{e\,i}\right)^2\, \frac{|p|^2}{M_{N_i}},
\label{effmassparanus3} 
\end{eqnarray}
with 
\begin{eqnarray}
&&{\large\bf M}_{\rm eff}={\large\bf M}^{ee}_{\nu} +{\large\bf
    M}^{ee}_{S}+{\large\bf M}^{ee}_{N}.\label{sumeff}.
\end{eqnarray}

The quantity ${\hat m}_{S_i}$ is the i-th eigen value of the  $S-$ fermion mass matrix
$m_S$. The magnitude of neutrino virtuality momentum $|p|$ has been estimated to be in the allowed range 
$|p|= 120$ MeV$-200$ MeV \cite{Vergados,Ver-a,Ver-b,Doi,Doi-a,Doi-b,Doi-c}.
The RH$\nu$s being much heavier than the singlet fermions, their
contributions have been neglected. 

\subsection{Singlet Fermion Assisted Enhanced Double Beta Decay Rate} 
We use neutrino oscillation data to estimate $M_{\nu}^{\rm ee}$ for NH and IH cases
with the values of Dirac phase and Majorana phases as discussed above. We further use the values of $M_i$ from Table \ref{tab:msMi} and Fig. \ref{fig:msM} and the Dirac neutrino mass matrix from eq.(\ref{eq:mdmu} to estimate $M_S^{\rm ee}$ 
while treating the RH$\nu$ mass at its assumed degenerate value of $M_{N_i}=4 {\rm TeV} (i=1,2,3)$. 
The variation of effective parameter $m_{\rm ee}$ as a function of
lightest neutrino mass is shown in Fig.\ref{fig:eff2} when $m_{s_1}=2$ GeV.
\begin{figure}[htbp]
 \includegraphics[width=9cm,height=9cm,angle=-90]{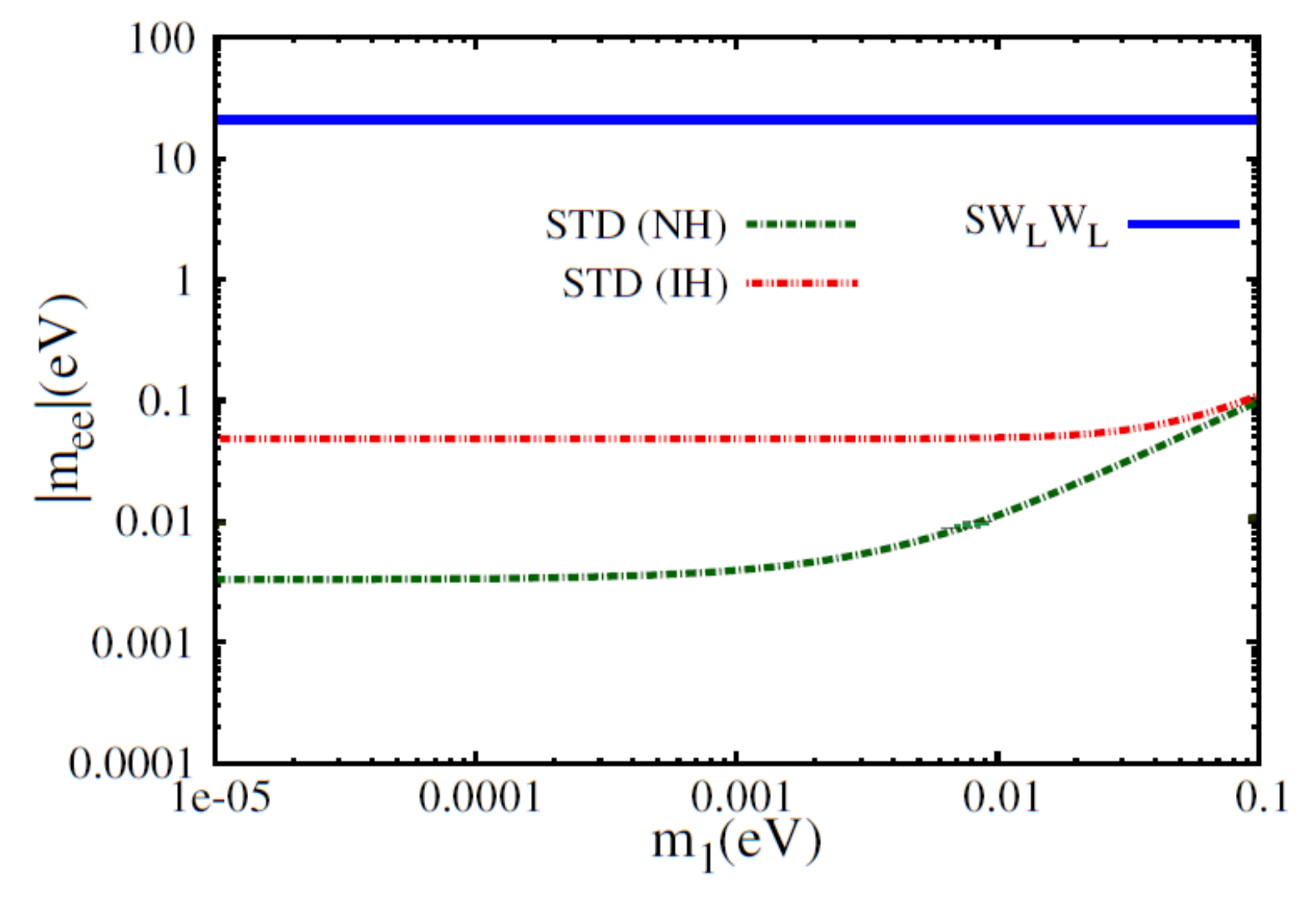}
 \caption{Variation of effective mass parameter as a function of
   lightest active neutrino mass $m_1$ for $m_{s_1}=2$ GeV. For
   comparison, predictions in the standard model suplementd by light neutrino masses
   of NH type is shown by green dot-dashed curve. For IH pattern of
   mass hierarchy the standard prediction is shown by red dot-dashed
   curve.}   
\label{fig:eff2}
\end{figure}

As noted from the analytic formulas, the effective mass parameter in
the singlet femion dominated case being inversely proportional to
$m_{s_1}$, it will proportionately decrease with increasing value of
the mediating particle mass. This feature has been shown in Fig.\ref{fig:eff4}.
\begin{figure}[htbp]
 \includegraphics[width=7cm,height=8cm,angle=-90]{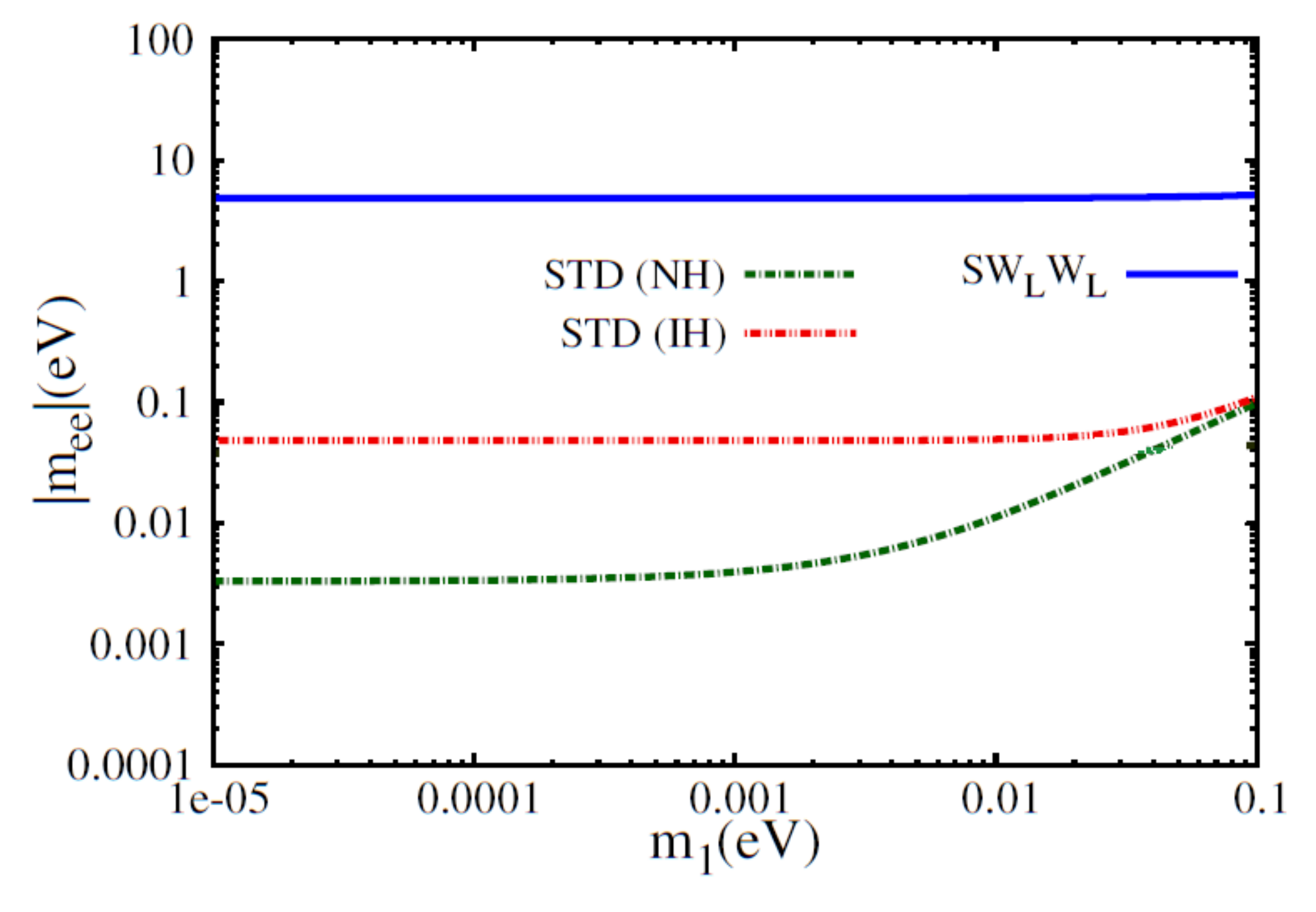}
 \caption{Same as Fig.\ref{fig:eff2} but for  $m_{s_1}=4.0$ GeV. }   
\label{fig:eff4}
\end{figure}
We present predictions of double-beta decay half life as a function of
the singlet fermion mass in Fig.\ref{fig:bbls1}. It is clear that while for
$m_{s_1}=2$ GeV the half life saturates the current experimental
limit, for larger values of $m_{s_1}$  the halflife is found to increase.
\begin{figure}[htbp]
 \includegraphics[width=9cm,height=12cm,angle=-90]{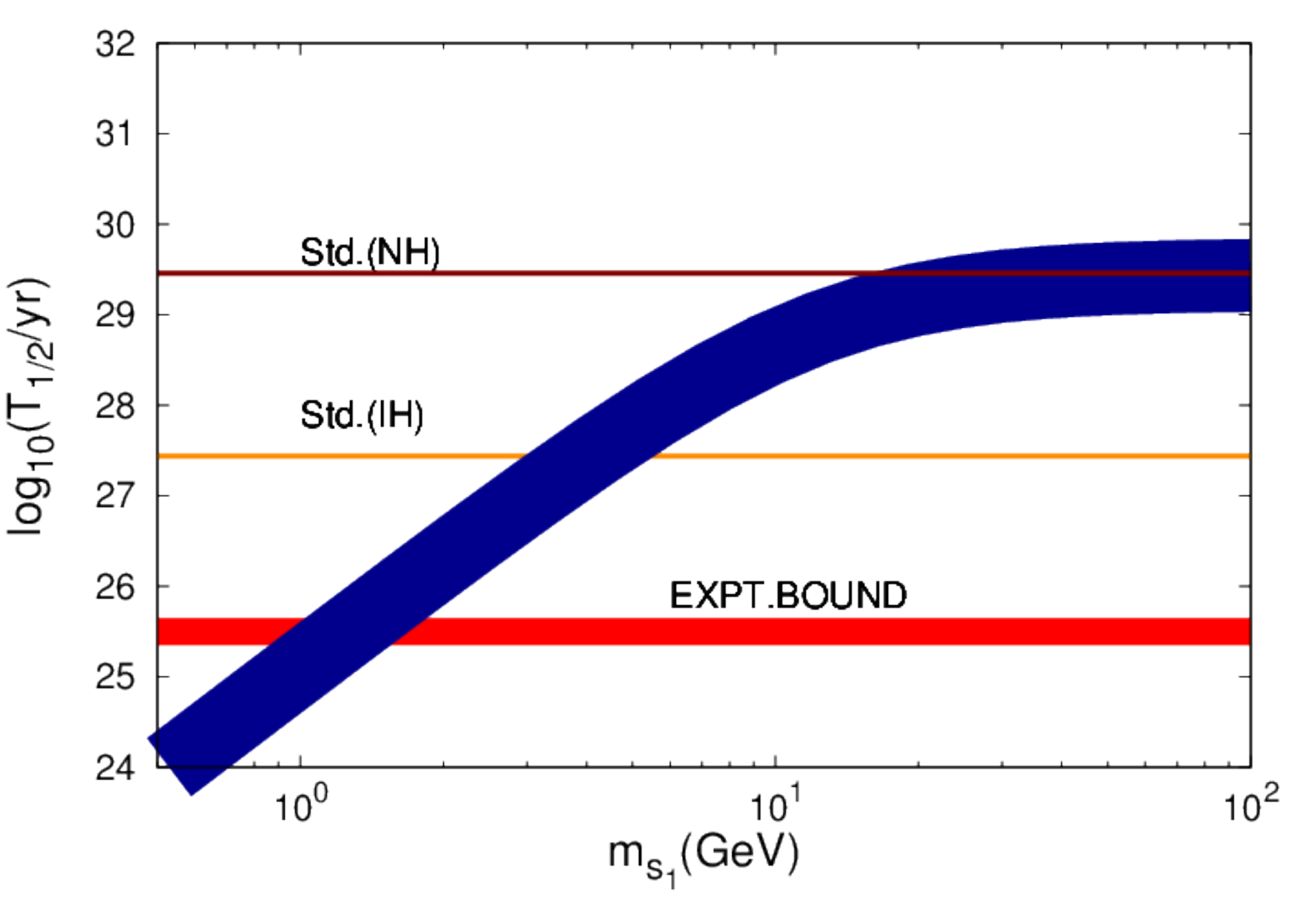}
 \caption{Prediction of double beta decay half-life 
 as a function of
   sterile neutrino mass  $m_{s_1}$ GeV (blue shaded curve) where the NH type  light
   neutrino and the sterile neutrino exchange contributions
   have been included. Effects of much larger masses $(m_{S_2}, m_{S_3})\gg
   m_{S_1}$ have been neglected. The spread in the curve reflects uncertainty in the virtuality momentum $p=120-190$ MeV. For comparison, the standard
   prediction with NH and IH pattern of light neutrino mass
   hierarchies are shown by the two respective horizontal lines. The
   bottom most thick red horizontal line closest to the X-axis represents 
    overlapping experimental bounds from different groups \cite{bbexpt1-Klapdoor,bbexpt2-a,bbexpt2-b,bbexpt3-a,bbexpt3-b}.  }   
\label{fig:bbls1}
\end{figure}
Neglecting heavy RH$\nu$ contributions but including those due to the lightest
sterile neutrino and the IH type light neutrinos our predictions of
half life as a function of the lightest sterile neutrino mass is shown
in Fig. \ref{fig:bbIH}  

\begin{figure}[htbp]
 \includegraphics[width=9cm,height=12cm,angle=-90]{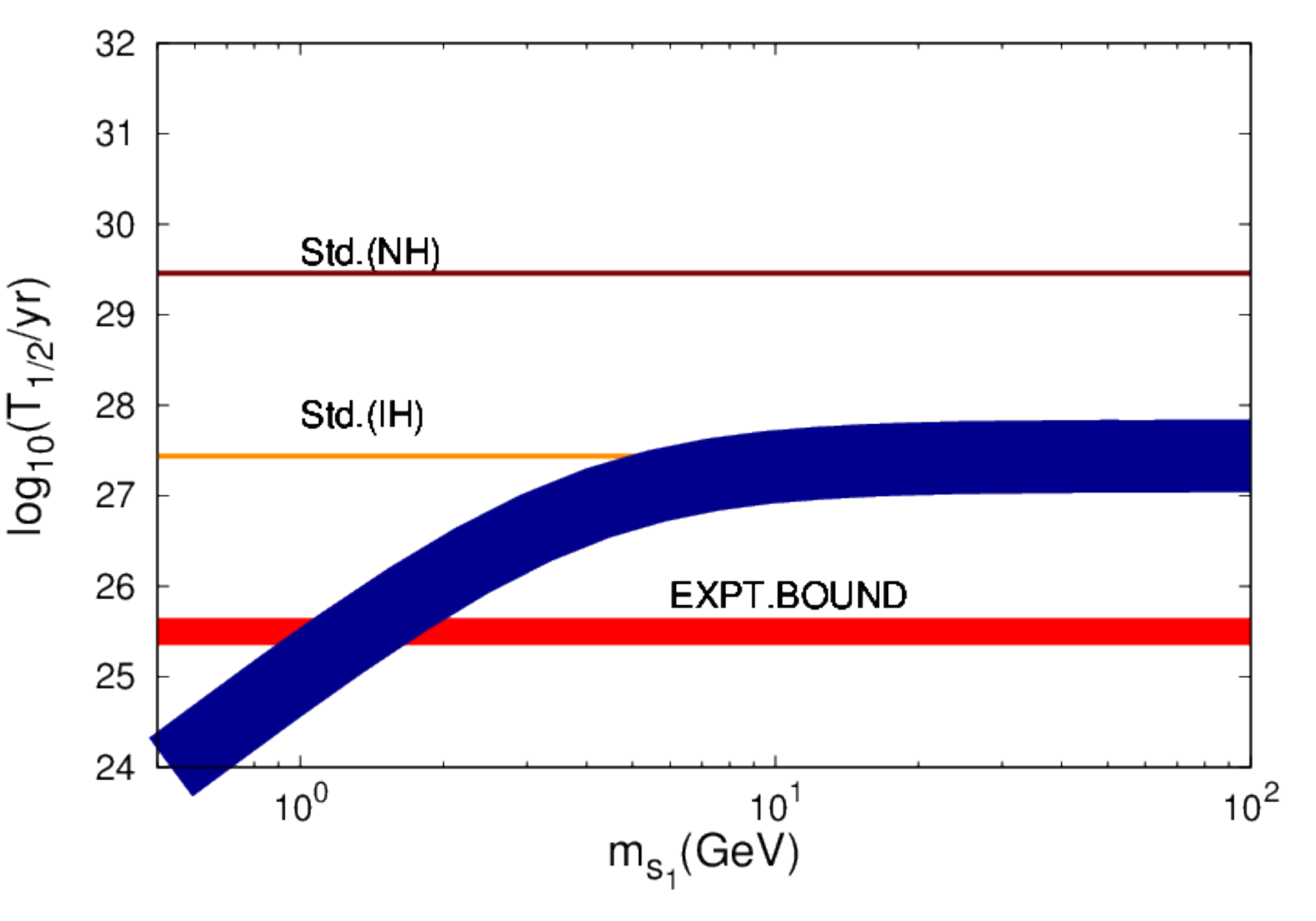}
 \caption{Same as Fig.\ref{fig:bbls1} but with contributions of IH type light neutrinos cambined with lightest sterile neutrinos.}
\label{fig:bbIH}
\end{figure}

Predicted lifetimes are seen to decrease with increasing sterile
neutrino mass. The sterile neutrino exchange contribution completely
dominates over light neutrino
exchange contributions for 
$m_{S_1}=1.3-7$ GeV in case of IH but for $m_{S_1}=1.5-20$ GeV in case of NH.  
At $m_{S_1}\simeq 1.5 GeV$ both types of solutions saturate the current laboratory limits reached by different experimental groups.

\section{SUMMARY, DISCUSSION AND CONCLUSION}\label{sec:sum}
A recently proposed scalar extension of minimal non-SUSY SU(5) GUT
has been found to realize  precision gauge coupling unification, high
scale type-II seesaw ansatz for neutrino masses, and prediction of  a
WIMP scalar DM candidate that also completes vacuum stability of the
scalar potential. But the LFV decays are predicted to have negligible
rates inaccessible to ongoing searches in foreseeable future. Like
wise experimentally verifiable double beta decay rates measurable by
different search experiments are possible only for quasi-degenerate
neutrino mass spectrum with large common mass scale $|m_0| > 0.2$ eV
or  $\sum_i m_i >
0.6$ eV. This
 violates the recently measured cosmological bound $\sum_i m_i \le
0.23$ eV. In order to remove these theoretical shortcomings in the
context of SU(5), we have extended this model
by the addition of three RH$\nu$s, three extra Majorana fermion
singlets $S_i(i=1,2,3)$ and a scalar singlet $\chi_S(1,0,1)$ that
generates $N-S$ mixing mass term through its vacuum expectation
value. In the original thory  of type-I seesaw cancellation mechanism,
although the choice of particles is same as $N_i$, $S_i$ and
$\chi_S(1,0,1)$,  the neutrino mass is given by double seesaw
\cite{Kim-Kang:2006}. Further, there is no grand unification of gauge couplings or prediction of proton decay in this model \cite{Kim-Kang:2006}, and the scalar potential of the model has vacuum instability . In addition the  
 $N_i$ are not gauged. The model does not  predict dominant contributions to double beta decay  for NH or IH type 
neutrino masses. In non-SUSY  SO(10) models of unification of three forces
 implementing  the cancellation of type-I seesaw \cite{mkp-bs:2015,app:2013,pas:2014}, the TeV scale RH neutrinos are gauged but the neutrino masses are controlled by inverse seesaw. But in \cite{nurev:mkpbpn,bpn-mkp:2015} the RH$\nu$s are gauged and the neutrino mass formula is linear seesaw or type-II seesaw \cite{nurev:mkpbpn}.  In all type-II seesaw
dominated  SO(10) models, the  RH$\nu$ masses have the same hierarchy
as the left-handed neutrino
masses:$M_{N_1}:M_{N_2}:M_{N_3}::m_1:m_2:m_3$. This happens precisely
because the left-handed and the right-handed dilepton Yukawa
interactions originate from the same SO(10) invariant term:
$f{16}_F{16}_F {126}^{\dagger}$. In SU(5), however, as the LH triplet $\Delta_L(3,-1,1)$
generating type-II seesaw and the  singlet $\sigma_S(1,0,1)$
generating RH$\nu$s  belong to different scalar representations,
${15}_H\subset SU(5)$ and ${\bf 1}_H\subset SU(5)$, respectively, they
can  possess different Majorana couplings in their respecive Yukawa interactions:$fll{\Delta_L}^C$ and
$f_N\sigma_S N N$. Because of this reason the the generated RH$\nu$
masses through $M_N=f_N\langle \sigma_s \rangle$ no longer follows the
predicted type-II seesaw predicted hierarchical pattern. Then the allowed fine tuning
$|f_N|<< |f|$ permits  RH neutrino mass scale $M_N\sim {\cal O}(1-10)$ TeV even though,
unlike SO(10) models, there are no possibilities of low mss $W_R$ or $Z^{\prime}$ bosons at this scale. The apprehension of unacceptably large active neutrino mass generation through type-I seesaw mechanism is rendered inoperative
through  the well established procedure of cancellation mechanism that is
also shown to operate profoundly in this SU(5) model. Such RH$\nu$s  generating $N-S$  mixing mass $M\simeq {\cal O} (100-1000)$ GeV now reproduce the well known results on
 LFV decay branching ratios only $4-5$ orders lower than the current
 experimental limit as well as the extensively investigated
 non-unitarity effects. Through the sterile neutrino canonical seesaw
 formula emerging from this cancellation mechanism (in the presence of
 $N_i$),  $m_S=-M\frac{1}{M_N}M^T$, this mechanism predicts their masses over a wide range of values, $m_{s_1}={\cal O} (1-100)$ GeV and $m_{s_2},m_{s_3} \sim {\cal O}(10-1000)$ GeV. The lightest sterile neutrino mass $m_{s_1}$ now predicts dominant double beta decay in the $W_L-W_L$ channel through the $\nu-S$ mixing close to the current experimental limits even though the light neutrino masses are of NH or IH type ($m_i << |0.2|$ eV) which satisfy the cosmological bound. For larger values of $m_{S_1}$ the predicted decay rate decreases  
and the sterile neutrino contribution becomes negligible for $m_{s_i}>> 50$ GeV.
In the limiting case when all the singlet fermion masses have such large values, the double beta decay rates asymptotically  approach the respective standard NH or IH type  contributions.   
The new mechanism of RH$\nu$ mass generation also allows the second
and the third generation sterile neutrino masses to be
quasi-degenerate (QD) near $1-10$ TeV scale while keeping $m_{S_1}\sim
1-10$ GeV suitable for dominant double beta decay mediation. There is a possibility that such TeV
scale QD masses while maintaining observable predictions on LFV decays  can effectively generate baryon asymetry of the universe
via resonant leptogenesis \cite{bpn-mkp:2015}.
 A scalar singlet DM can be easily
accommodated as discussed in \cite{scp:2018,falkowski:2014,spc:2017} while resolving the issue
of vacuum stability. Irrespective of scalar DM,
the model can also  accommodate a Majorana fermion singlet dark matter \cite{strumia:2006} which can emerge from the additional fermionic representation ${24}_F\subset SU(5)$ .         

The prediction of new fermions has an additional advantage over scalars as 
these masses are protected by leptonic global symmetries
\cite{tHooft}. Also
the prediction of such Majorana type sterile neutrinos  can be tested
by high enegy and high luminousity accelerators through their
like-sign dilepton production processes  \cite{bpn-mkp:arx}. For example at LHC they can mediate the process $pp\to W_L X \to l^{\pm}l^{\pm}jjX$ where the jets could manifest as mesons. It would be quite interesting to examine emergence of such SU(5) theory as a remnant of SO(10) or $E_6$ GUTs. 

We conclude that even in the presence of SM as effective gauge theory
  descending from a suitable SU(5) extension, it is possible to predict
  experimentally accessible double beta decay rates in the $W_L-W_L$ channel satisfying the
  cosmological bound on active neutrino masses as well as verfiable
  LFV decays. The RH$\nu$ masses can be considerably
  different from those constrained by conventional type-I or type-II
  seesaw frameworks which are instrumental in predicting interesting physical phenomena even
  if there are no non-standard heavy gauge bosons anywhere below the 
GUT scale.

------------------------------------------------------------\\
\appendix
\section{APPENDIX:\,Diagonalisation, Masses, and Mixings}
\label{sec:app-massmix}
The purpose of this Appendix is to certain details of mixings among
the fermions $\nu$, $S$, and $N$ and also derive their masses by block
diagonalisation of the resulting $9\times 9$ neutral fermion mass
matrix discussed in Sec.\ref{sec:cancel}. 
 We write the complete  $9\times 9$ mass matrix
   in the 
flavor basis $\{\nu_L, S_L, N^C_R\}$ after the effect of $\Delta_L$
is integrated out 
\bea
\mathcal{M}_\nu=
\bmt
m^{(II)}     &  0      &  M_D  \\
0     &  0  &  M^T  \\
M_D^T &  M      &  M_N
\emt
\label{app:numass}
\eea
where the type-II seesaw contribution has been noted as $m^{(II)}=fv_L$.
The flavor basis to mass basis transformation and  diagonalisation of  $\mathcal{M}_\nu$
is achieved by  a unitary transformation matrix ${\cal V}$ defined below
\bea 
& &|\psi\rangle_f=\mathcal{V}\, |\psi\rangle_m \\
&\mbox{or,}&\,\bmt 
\nu_\alpha\\ S_\beta \\ N^C_\gamma
\emt
=
\bmt 
{\cal V}^{\nu\nu}_{\alpha i} & {\cal V}^{\nu{S}}_{\alpha j} & {\cal V}^{\nu {N}}_{\alpha k} \\
{\cal V}^{S\nu}_{\beta i} & {\cal V}^{SS}_{\beta j} & {\cal V}^{SN}_{\beta k} \\
{\cal V}^{N\nu}_{\gamma i} & {\cal V}^{NS}_{\gamma j} & {\cal V}^{NN}_{\gamma k} 
\emt
\bmt 
\hat{\nu}_i \\ \hat{S}_j \\ \hat{N}_k
\emt  
 \label{app:formmix}
\\
&\mbox{and} &\,\mathcal{V}^\dagger \mathcal{M}_\nu \mathcal{V}^*
    =  \hat{\mathcal{M}}_\nu
	 = {\rm diag}\left({ \hat{m}}_{\nu_i};{ \hat{m}}_{{\cal S}_j};{ \hat{m}}_{{\cal N}_k}\right)
	 \label{app:massdiag}
\eea
where subscripts $f, m$ denote for the flavor and mass basis, respectively. Also $\mathcal{M}_\nu$ 
is the mass matrix in flavor basis with $\alpha, \beta, \gamma$
running  over three generations of light-neutrinos, 
sterile-neutrinos and right handed heavy-neutrinos. Here $\hat{\mathcal{M}}_\nu$ is 
the diagonal mass matrix with $(i,j,k=1,2,3)$ running over corresponding mass states at the sub-eV, GeV 
and TeV scales, respectively. 

 The method of complete diagonalization will be carried 
out in two steps: ({\bf 1}) the full neutrino mass matrix $\mathcal{M}_\nu$ has to reduced to a block 
diagonalized form as $\mathcal{M}_{\rm \tiny BD}$, ({\bf 2}) this block diagonal form further diagonalized 
to give physical masses of the neutral leptons $\hat{\mathcal{M}}_{\nu}$. 

\par\noindent{\bf (1) Determination of $\mathcal{M}_{\rm \tiny BD}$}\\
We shall follow the parametrisation of the type given in
Ref. \cite{LG:2000} to determine the form of the diagonalising matrices
${\mathcal{W}}_1$ and ${\mathcal{W}}_2$. We define their product as  
\be
\mathcal{W}=\mathcal{W}_1\, \mathcal{W}_2 \label{eq:w}
\ee
 where $\mathcal{W}_{1}$ and $\mathcal{W}_2$ satisfy 
\bea 
\mathcal{W}_1^\dagger \mathcal{M}_\nu  \mathcal{W}^*_1 = \hat{\mathcal{M}}_{\rm \tiny BD}, \mbox{and}\quad 
\mathcal{W}_2^\dagger \hat{\mathcal{M}}_{\rm \tiny BD}  \mathcal{W}^*_2 = \mathcal{M}_{\rm \tiny BD}
\eea
Here $\hat{\mathcal{M}}_{\rm \tiny BD}$, and $\mathcal{M}_{\rm \tiny BD}$ are the intermediate block-diagonal, 
and full block-diagonal mass matrices, respectively,
\bea 
& &\hat{\mathcal{M}}_{\rm \tiny BD} =
\bmt
{\cal M}_{eff}&0\\
0& m_{\cal N}
\emt \\
&\mbox{and}&\, \mathcal{M}_{\rm \tiny BD}
= \bmt m_\nu&0&0\\
0&m_{\cal S}&0\\
0&0&m_{\cal N}
\emt
\eea
\par\noindent{(2)\bf {Determination of $\mathcal{W}_1$}}\\
We need to first integrate out the heavy state ($N_R$), being heavier than other mass scales 
in our theory, such that up to the leading order approximation the analytic expressions for 
$\mathcal{W}_1$ is
\bea 
\mathcal{W}_1=\bmt
1-\frac{1}{2}B^*B^T&B^*\\
-B^T&1-\frac{1}{2}B^TB^*
\emt\,, 
\eea 
where the matrix $B$ is $6\times 3$ dimensional and is described as
 
\be 
B^\dagger =M_N^{-1}\left(M^T_D, M^T\right)=(Z^T, Y^T)
\ee
where, $X=M_DM^{-1}$, $Y=M{M_N}^{-1}$, and $Z=M_DM_N^{-1}$ so that 
$Z=X \cdot Y\neq Y\cdot X$ . 

Therefore, the transformation matrix $\mathcal{W}_1$ can be written purely in terms of dimensionless parameters $Y$ and $Z$
\bea 
\mathcal{W}_1=\bmt
1-\frac{1}{2}ZZ^\dagger & -\frac{1}{2}ZY^\dagger & Z \\
-\frac{1}{2}YZ^\dagger & 1-\frac{1}{2}YY^\dagger & Y \\
-Z^\dagger & -Y^\dagger & 1-\frac{1}{2}(Z^\dagger Z + Y^\dagger Y)
\emt
 \label{app:w1}
\eea
while the light and heavy states can be now written as
\bea
{\cal M}_{eff}&=&-
\bmt 
M_DM_N^{-1}M^T_D&M_DM_N^{-1}M^T\\
MM_N^{-1}M^T_D& MM_N^{-1}M^T
\emt \\
{ m}_{\cal N}&=&M_N+..
\eea
\par\noindent{Determination of $\mathcal{W}_2$}\\
From the above discussion, it is quite clear now that the eigenstates $\mathcal{N}_i$ are 
eventually decoupled from others and the remaining mass matrix ${\cal M}_{eff}$ can be 
block diagonalized using another transformation matrix
\bea 
\mathcal{S}^\dagger \mathcal{M}_{\rm eff} \mathcal{S}^* 
      = \bmt 
         m_\nu & 0 \\
         0     & m_{\cal S}
        \emt
\eea
such that
\bea
\mathcal{W}_2 =  \bmt
               \mathcal{S}&0\\
               0&1
                 \emt 
\eea
In a simplified structure 
\bea 
{\cal M}_{eff}=\bmt 
m^{II}_\nu +M_DZ^T & M_DY^T \\
YM_D^T & MY^T
\emt
\eea
Under the assumption at the  beginning $Z<<Y$, and of-course $M_D<<M$, this structure is similar to type-(I+II) seesaw. 
Therefore we immediately get the light neutrino masses as
\bea 
m_\nu&=&-M_DZ^T+m^{II}_{\nu}+M_DY^T(MY^T)^{-1}YM^T\nn\\
&=&-M_DZ^T+M_DZ^T+m^{II}_{\nu}         \nn\\
&=&m^{II}_{\nu} \\
m_{\cal S}&=&-MM_N^{-1}M^T
\eea
We see that in addition to $m_{\cal N}$ the $m_{\cal S}$ is also almost diagonal if $M$ and $M_N$ 
are assumed to be diagonal. The transformation matrix $S$ is 

\bea 
S=\bmt 
1-\frac{1}{2} A^*A^T&A^*\\
-A^T&1-\frac{1}{2}A^TA^*
\emt
\eea
such that
\bea 
A^\dagger&=&(MY^T)^{-1}YM_D^T\nn\\
&\simeq& (MY^T)^{-1}YM_D^T=X^T\, .
\eea
The $3 \times 3$ block diagonal mixing matrix $\mathcal{W}_2$ has the following form
\bea 
\mathcal{W}_2 
=\bmt 
S & {\bf 0} \\
{\bf 0} & {\bf 1}
\emt = 
\bmt 
1-\frac{1}{2}XX^\dagger &X & 0\\
-X^\dagger & 1-\frac{1}{2}X^\dagger X & 0 \\
0 & 0 & 1
\emt
 \label{app:w2}
\eea
\subsection{Physical Neutrino Masses from Complete Diagonalization }
The block diagonal matrices $m_\nu$, $m_{\cal S}$ and $m_{\cal N}$ can further be diagonalized 
to give physical masses for all neutral leptons by a unitary matrix $\mathcal{U}$ as
\bea
\mathcal{U}=\bmt U_\nu & 0 & 0 \cr 0 & U_S & 0 \cr 0 & 0 & U_N \emt.
\label{eq:mixb}
\eea
where the unitary matrices $U_\nu$, $U_{S}$ and $U_{N}$ satisfy
\begin{eqnarray}
U^\dagger_\nu\, m_{\cal\nu}\, U^*_{\nu}  &=& \hat{m}_\nu = 
         \text{diag}\left(m_{\nu_1}, m_{\nu_2}, m_{\nu_3}\right)\, , \nonumber \\ 
U^\dagger_S\, m_{\cal S}\, U^*_{S}  &=& \hat{m}_S = 
         \text{diag}\left(m_{S_1}, m_{S_2}, m_{S_3}\right)\, , \nonumber \\
U^\dagger_N\, m_{\cal N}\, U^*_{N}  &=& \hat{m}_N = 
         \text{diag}\left(m_{N_1}, m_{N_2}, m_{N_3}\right)\,
 \label{app:unit}
\end{eqnarray}
\noindent
With this discussion, the complete mixing matrix is
 \begin{eqnarray}
\mathcal{V}&=&\mathcal{W} \cdot \mathcal{U} =
\mathcal{W}_{1}\cdot \mathcal{W}_{2}\cdot \mathcal{U} \nonumber \\
&=&
\bmt
1-\frac{1}{2}ZZ^\dagger & -\frac{1}{2}ZY^\dagger & Z \\
-\frac{1}{2}YZ^\dagger & 1-\frac{1}{2}YY^\dagger & Y \\
-Z^\dagger & -Y^\dagger & 1-\frac{1}{2}(Z^\dagger Z + Y^\dagger Y)
\emt
\bmt
1-\frac{1}{2}XX^\dagger & X & 0 \\
-X^\dagger  &   1-\frac{1}{2}X^\dagger X & 0 \\
0 & 0 & 1
\emt
\bmt 
U_\nu &0&0\\
0&U_{S}&0\\
0&0&U_{N}
\emt  \nonumber \\
&=&
\bmt
1-\frac{1}{2}XX^\dagger & X-\frac{1}{2}ZY^\dagger & Z \\
-X^\dagger & 1-\frac{1}{2}(X^\dagger X+YY^\dagger) & Y-\frac{1}{2}X^\dagger Z\\
0&-Y^\dagger&1-\frac{1}{2}Y^\dagger Y
\emt
\cdot
\bmt 
U_\nu &0&0\\
0&U_{S}&0\\
0&0&U_{N}
\emt 
 \label{app:mix-extended}
 \end{eqnarray}
It is straight forward to verify that  the matrix product of the right-hand side of
eq.(\ref{app:mix-extended})  agrees with eq.(\ref{eqn:Vmix-extended})
of Sec.\ref{sec:cancel}.
\section {\bf ACKNOWLEDGMENT }
M. K. P. thanks the Science and Engineering Research Board, Department
of Science and Technology, Government of India for grant of research
project SB/S2/HEP-011/2013. R.S. thanks  Siksha 'O' Anusandhan 
University for research fellowship.\\


\end{document}